\documentclass{acm_proc_article-sp}
\usepackage{amssymb}
\usepackage{courier,cite}
\usepackage{flushend}

\usepackage{listings}
\usepackage{placeins}
\usepackage{subcaption}
\usepackage{tikz}

\newcommand{\mpi}[1]{\textsf{MPI\_#1}}
\newcommand{\fompi}{\textsc{foMPI}}

\DeclareCaptionFont{white}{\color{white}}
\DeclareCaptionFormat{listing}{\colorbox[cmyk]{0.43, 0.35, 0.35,0.01}{\parbox{\textwidth}{\hspace{15pt}#1#2#3}}}
\makeatletter
\def\@copyrightspace{\relax}
\makeatother

\begin{document}

\lstset{
         basicstyle=\footnotesize\ttfamily,
         numberstyle=\tiny,
         numbersep=5pt,
         tabsize=2,
         extendedchars=true,
         breaklines=true,
         keywordstyle=\color{red},
         frame=b,         
         stringstyle=\color{white}\ttfamily,
         showspaces=false,
         showtabs=false,
         xleftmargin=17pt,
         framexleftmargin=17pt,
         framexrightmargin=5pt,
         framexbottommargin=4pt,
         showstringspaces=false
}

\title{Enabling Highly-Scalable Remote Memory Access Programming with MPI-3 One Sided}

\numberofauthors{3}

\author{
\alignauthor
Robert Gerstenberger
\titlenote{The author performed much of the implementation during an 
internship at UIUC/NCSA while the analysis and documentation was 
performed during a scientific visit at ETH Zurich. The author's primary
email address is \\ gerstenberger.robert@gmail.com.}\\
  \affaddr{ETH Zurich}\\
  \affaddr{Dept. of Computer Science}\\
  \affaddr{Universit\"atstr. 6}\\
  \affaddr{8092 Zurich, Switzerland}
  \email{robertge@inf.ethz.ch}
\alignauthor
Maciej Besta\\
  \affaddr{ETH Zurich}\\
  \affaddr{Dept. of Computer Science}\\
  \affaddr{Universit\"atstr. 6}\\
  \affaddr{8092 Zurich, Switzerland}
  \email{maciej.besta@inf.ethz.ch}
\alignauthor
Torsten Hoefler\\
  \affaddr{ETH Zurich}\\
  \affaddr{Dept. of Computer Science}\\
  \affaddr{Universit\"atstr. 6}\\
  \affaddr{8092 Zurich, Switzerland}
  \email{htor@inf.ethz.ch}
}

\date{13 August 2013}

\maketitle

\begin{abstract}
Modern interconnects offer remote direct memory access (RDMA) features.
Yet, most applications rely on explicit message passing for
communications albeit their unwanted overheads. The MPI-3.0 standard
defines a programming interface for exploiting RDMA networks directly,
however, it's scalability and practicability has to be demonstrated in
practice. 
In this work, we develop scalable bufferless protocols that implement
the MPI-3.0 specification. Our protocols support scaling to millions of
cores with negligible memory consumption while providing highest
performance and minimal overheads. 
To arm programmers, we provide a spectrum of performance models for all
critical functions and demonstrate the usability of our library and
models with several application studies with up to half a million
processes.
We show that our design is comparable to, or better than UPC and Fortran
Coarrays in terms of latency, bandwidth, and message rate. We also
demonstrate application performance improvements with comparable
programming complexity.

\end{abstract}

\section{Motivation}

Network interfaces evolve rapidly to implement a growing set of features
directly in hardware. A key feature of today's high-performance networks
is remote direct memory access (RDMA). RDMA enables a process to
directly access memory on remote processes without involvement of the
operating system or activities at the remote side. This hardware support
enables a powerful programming mode similar to shared memory
programming.
RDMA is supported by on-chip networks in, e.g., Intel's SCC and IBM's
Cell systems, as well as off-chip networks such as
InfiniBand~\cite{IBAspec,ofed:url}, IBM's PERCS~\cite{ibm-percs-network} or
BlueGene/Q~\cite{PAMI:BlueGeneQ}, Cray's
Gemini~\cite{5577317} and Aries~\cite{Faanes:2012:CCS:2388996.2389136}, or even
RoCE over Ethernet~\cite{Beck:2011:PER:2043535.2043537}. 

\sloppy
From a programmer's perspective, parallel programming schemes can be
split into three categories: (1) shared memory with implicit
communication and explicit synchronization, (2) message passing with
explicit communication and implicit synchronization (as side-effect of
communication), and (3) remote memory access and partitioned global
address space (PGAS) where synchronization and communication are managed
independently. 

Architects realized early that shared memory can often not be
efficiently emulated on distributed
machines~\cite{Karlsson:1998:CCC:646092.680546}. Thus, message passing
became the \emph{de facto} standard for large-scale parallel
programs~\cite{mpi-3.0}. However, with the advent of RDMA networks, it
became clear that message passing over RDMA incurs additional overheads
in comparison with native remote memory access (RMA, aka. PGAS) 
programming~\cite{Nishtala:2009:SCA:1586640.1587648,Bell:2006:OBL:1898953.1899016,Bell:2003:ECH:838237.838390}.
This is mainly due to message matching, practical issues with
overlap, and because fast message passing libraries over RDMA usually
require different protocols~\cite{Woodall06highperformance}: an eager
protocol with receiver-side buffering of small messages and a rendezvous
protocol that synchronizes the sender. Eager requires additional copies,
and rendezvous sends additional messages and may delay the sending
process.  

In summary, directly programming RDMA hardware has benefits in the
following three dimensions: (1) \emph{time} by avoiding message matching
and synchronization overheads, (2) \emph{energy} by reducing
data-movement, e.g., it avoids additional copies of eager messages, and
(3) \emph{space} by removing the need for receiver buffering.
\begin{figure*}
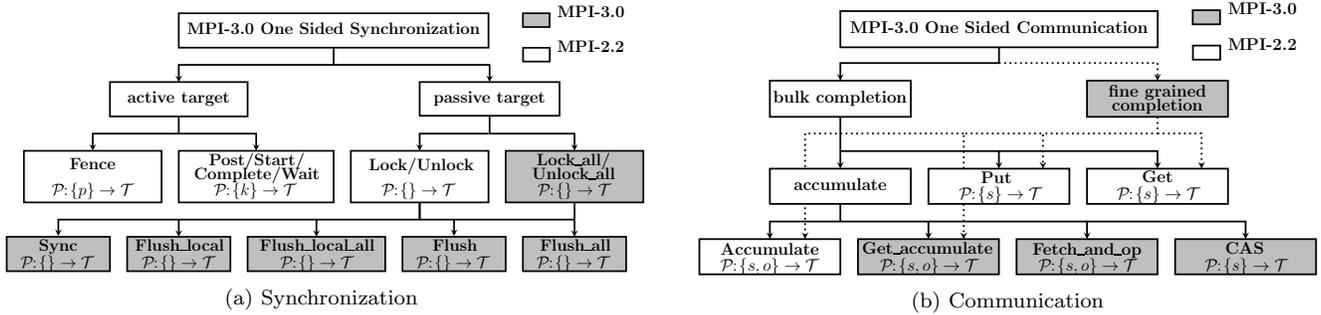

\centering
\begin{subfigure}{0.495\textwidth}
  \resizebox{\textwidth}{!}{
  \tikzset{labe/.style={\font={\huge\itshape}}}
    \input{img/overview_synch.tex}
  }
  \caption{Synchronization}
  \label{fig:sync_overview}
\end{subfigure}
\hfill
\begin{subfigure}{0.475\textwidth}
  \resizebox{\textwidth}{!}{
    \input{img/overview_comm.tex}
  }
  \caption{Communication}
  \label{fig:comm_overview}
\end{subfigure}
\caption{Overview of MPI-3.0 One Sided and associated cost functions.
The figure shows abstract cost
functions for all operations in terms of their input domains.
The symbol $p$ denotes the number of processes, $s$ is the data size,
$k$ is the maximum number of neighbors, and $o$ defines an MPI
operation. The notation $\mathcal{P}\!\!:\!\{p\}\rightarrow\mathcal{T}$
defines the input space for the performance (cost) function $\mathcal{P}$. In
this case, it indicates, for a specific MPI function, that the execution
time depends only on $p$. We provide asymptotic cost functions in
Section~\ref{sec:implem} and parametrized cost functions for our
implementation in Section~\ref{sec:micro}.}
\label{fig:overview}
\end{figure*}
Thus, several programming environments allow to access RDMA hardware
more or less directly: PGAS languages such as Unified Parallel C
(UPC~\cite{upc}) or Fortran Coarrays~\cite{fortran2008} and libraries such
as Cray SHMEM~\cite{shmem} or MPI-2.2 One Sided~\cite{mpi-2.2}. A lot of
experience with these models has been gained in the past
years~\cite{Nishtala:2009:SCA:1586640.1587648,Bell:2006:OBL:1898953.1899016,Zhang:2011:OBA:2063384.2063485}
and several key design principles for remote memory access (RMA)
programming evolved.
The MPI Forum set out to define a portable library interface to RMA
programming using these established principles. This new interface in
MPI-3.0~\cite{mpi-3.0} extends MPI-2.2's One Sided chapter to support
the newest generation of RDMA hardware. 

However, the MPI standard only defines a programming interface and does
not mandate an implementation. Thus, it has yet to be demonstrated that
the new library interface delivers comparable performance to compiled
languages like UPC and Fortran Coarrays and is able to scale to large
process numbers with small memory overheads. 
\emph{In this work, we develop scalable protocols for implementing
MPI-3.0 RMA over RDMA networks. We demonstrate that (1) the performance
of a library implementation can be competitive to tuned, vendor-specific
compiled languages and (2) the interface can be implemented on
highly-scalable machines with negligible memory overheads. In a wider
sense, our work answers the question if the MPI-3.0 RMA interface is a
viable candidate for moving into the post-petascale era.} 

Our key contributions are:
\begin{itemize}
  \item We describe scalable protocols and a complete implementation for 
    the novel MPI-3.0 RMA programming interface requiring 
    $\mathcal{O}(\log p)$ time and space per process on $p$ processes.
  \item We provide a detailed performance evaluation and performance
    models that can be used for algorithm development and to demonstrate
    the scalability to future systems.
  \item We demonstrate the benefits of RMA programming for several
    motifs and real-world applications on a multi-petaflop machine with
    full-application speedup of more than 13\% over MPI-1 using more than half a
    million MPI processes.
\end{itemize}

\section{Scalable Protocols for MPI-3.0 One Sided over RDMA Networks}
\label{sec:implem}

We describe protocols to implement MPI-3.0 One Sided
purely based on low-level remote direct memory access (RDMA). In all
our protocols, we assume that we only have small bounded buffer space at
each process, no remote software agent, and only put, get, and some
basic atomic operations for remote memory access. This makes our
protocols applicable to all current RDMA networks and is also
forward-looking towards exascale interconnect architectures. 

MPI-3.0 offers a plethora of functions with different performance
expectations and use-cases. We divide the RMA functionality 
into three separate concepts: (1) window creation, (2) communication
functions, and (3) synchronization functions.
In addition, MPI-3.0 specifies two memory models: a weaker model, called
``separate'', to retain portability and a stronger model, called
``unified'', for highest performance. In this work, we only consider the
stronger unified model since is it supported by all current RDMA
networks. More details on memory models can be found in the MPI
standard~\cite{mpi-3.0}.

Figure~\ref{fig:sync_overview} shows an overview of MPI's
synchronization functions. They can be split into active target mode, in
which the target process participates in the synchronization, and
passive target mode, in which the target process is passive.
Figure~\ref{fig:comm_overview} shows a similar overview of MPI's
communication functions. Several functions can be completed in bulk with
bulk synchronization operations or using fine-grained request objects
and test/wait functions.  However, we observed that the completion model only minimally
affects local overheads and is thus not considered separately in the
remainder of this work. 

Figure~\ref{fig:overview} also shows abstract definitions of
the performance models for each synchronization and communication
operation. The precise performance model for each function depends on
the exact implementation. 
We provide a detailed overview of the \emph{asymptotic} as well as
\emph{exact} performance properties of our protocols and our
implementation in the next sections. 
The different performance characteristics of communication and
synchronization functions make a unique combination of implementation
options for each specific use-case optimal. However, it is not always
easy to choose this best variant.  
The exact models can be used to design such optimal implementations (or
as input for model-guided
autotuning~\cite{Fraguela:2009:ATD:1636712.1637765}) while the simpler
asymptotic models can be used in the algorithm design phase
(cf.~\cite{Karp:1993:OBS:165231.165250}).

To support post-petascale computers, all protocols need to implement
each function in a scalable way, i.e., consuming $\mathcal{O}(\log p)$
memory and time on $p$ processes.
For the purpose of explanation and illustration, we choose to discuss a
reference implementation as use-case. However, all protocols and schemes
discussed in the following can be used on any RDMA-capable network.

\subsection{Use-Case: Cray DMAPP and XPMEM}

We introduce our implementation \fompi{} (fast one sided
MPI), a fully-functional MPI-3.0 RMA library implementation for Cray 
Gemini (XK5, XE6) and Aries (XC30) systems.
In order to maximize asynchronous progression and minimize overhead,
\fompi{} interfaces to the lowest available hardware APIs. 

For inter-node (network) communication, \fompi{}
\footnote{\fompi{} can be downloaded from \\
\small{http://spcl.inf.ethz.ch/Research/Parallel\_Programming/foMPI}}
uses the lowest-level networking
API of Gemini and Aries networks, DMAPP (Distributed Memory
Application), which has direct access to the hardware (GHAL) layer. 
DMAPP provides an RDMA interface and each process can expose (register)
local memory regions. Accessing remote memory requires a special key
(which is returned by the registration call).
DMAPP offers put, get, and a limited set of atomic memory operations, each of them comes
in three categories: blocking, explicit nonblocking, and implicit
nonblocking. All explicit nonblocking routines return a handle that can
be used to complete single operations, implicit nonblocking operations
can only be finished by bulk completion (gsync) functions.
DMAPP put and get can operate on 1, 4, 8 and 16 Byte chunks while atomic memory
operations (AMO) always operate on 8 Bytes.

For intra-node communication, we use XPMEM\cite{xpmem-white}, a portable
Linux kernel module that allows to map the memory of one process into
the virtual address space of another. Similar to DMAPP, processes can
expose contiguous memory regions and other processes can attach (map)
exposed regions into their own address space. All operations can then be
directly implemented with load and store instructions, as well as CPU
atomics (e.g., using the x86 lock prefix). Since XPMEM allows direct
access to other processes' memory, we include it in the category of RDMA
interfaces.

\fompi{}'s performance properties are
self-consistent~\cite{LarssonTraff:2010:SMP:1803941.1804141} and thus
avoid surprises for users.
We now proceed to develop algorithms to implement the window creation
routines that expose local memory for remote access. After this, we
describe protocols for synchronization and communication functions over
RDMA networks. 

\subsection{Scalable Window Creation}

A \emph{window} is a region of process memory that is made accessible to
remote processes. MPI-3.0 provides four collective functions for
creating different types of windows: \mpi{Win\_create} (traditional
windows), \mpi{Win\_allocate} (allocated windows),
\mpi{Win\_create\_dynamic} (dynamic windows), and
\mpi{Win\_allocate\_shared} (shared windows). We assume that
communication memory needs to be registered with the communication
subsystem and that remote processes require a remote descriptor that is
returned from the registration to access the memory. This is true for
most of today's RDMA interfaces including DMAPP and XPMEM. 

\paragraph{Traditional Windows} These windows expose existing
user-memory to remote processes. 
Each process can specify an arbitrary local base address for the window
and all remote accesses are relative to this address. This essentially
forces an implementation to store all remote addresses separately. This
storage may be compressed if addresses are identical, however, it
requires $\Omega(p)$ storage on each of the $p$ processes in the worst
case. 

Each process discovers intra-node and inter-node neighbors and
registers the local memory using XPMEM and DMAPP. Memory descriptors and
various information (window size, displacement units, base pointer, et
cetera) can be communicated with two \mpi{Allgather} operations: the
first with all processes of the window to exchange DMAPP information and
the second with the intra-node processes to exchange XPMEM information. 
Since traditional windows are fundamentally non-scalable, and only
included in MPI-3.0 for backwards-compatibility, their use is strongly
discouraged. 

\paragraph{Allocated Windows} Allow the MPI library to allocate window
memory and thus use a symmetric heap, where the base addresses on all
nodes are the same requiring only $\mathcal{O}(1)$ storage.
This can either be done by allocating windows in a system-wide symmetric
heap or with the following POSIX-compliant protocol: (1) a leader
(typically process zero) chooses a random address which it broadcasts to
all processes in the window, and (2) each process tries to allocate the
memory with this specific address using \verb+mmap()+. Those two steps
are repeated until the allocation was successful on all processes (this
can be checked with \mpi{Allreduce}). Size and displacement unit can now
be stored locally at each process. This mechanism requires
$\mathcal{O}(1)$ memory and $\mathcal{O}(\log p)$ time (with high
probability).

\paragraph{Dynamic Windows} These windows allow the dynamic
attach and detach of memory regions using \mpi{Win\_attach} and
\mpi{Win\_detach}.
Attach and detach operations are non-collective. In our implementation,
attach registers the memory region and inserts the information into a
linked list and detach removes the region from the list. Both operations
require $\mathcal{O}(1)$ memory per region.

The access of the list of memory regions on a target is purely one sided
using a local cache of remote descriptors. Each process maintains an
\verb~id~ counter, which will be increased in case of an attach or
detach operation. A process that attempts to communicate with the
target first reads the \verb~id~ (with a get operation) to check if its
cached information is still valid. If so, it finds the remote descriptor
in its local list. If not, the cached information are discarded, the
remote list is fetched with a series of remote operations and stored in
the local cache.

Optimizations can be done similar to other distributed cache protocols.
For example, instead of the \verb~id~ counter, each process could
maintain a list of processes that have a cached copy of its local
memory descriptors. Before returning from detach, a process notifies all
these processes to invalidate their cache and discards the remote
process list. For each communication attempt, a process has to first
check if the local cache has been invalidated (in which case it will be
reloaded).  Then the local cache is queried for the remote descriptor.
If the descriptor is missing, there has been an attach at the target and
the remote descriptor is fetched into the local cache.  After a cache
invalidation or a first time access, a process has to register itself on
the target for detach notifications. We explain a scalable data structure
that can be used for the remote process list in the General Active
Target Synchronization part (see Figure~\ref{fig:pscw_example:expl}).

The optimized variant enables better latency for communication
functions, but has a small memory overhead and is suboptimal for
frequent detach operations.

\paragraph{Shared Memory Windows}

Shared memory windows can be implemented using POSIX shared memory or
XPMEM as described in~\cite{mpi-shared-mem-win} with constant memory
overhead per core. Performance is identical to our direct-mapped (XPMEM)
implementation and all operations are compatible with shared memory
windows.

We now show novel protocols to implement synchronization modes in a
scalable way on pure RDMA networks without remote buffering.

\subsection{Scalable Window Synchronization}
MPI differentiates between \emph{exposure} and \emph{access} epochs. A
process starts an exposure epoch to allow other processes to access its
memory. In order to access exposed memory at a remote target, the origin process
has to be in an access epoch. Processes can be in access and exposure
epochs simultaneously and exposure epochs are only defined for active
target synchronization (in passive target, window memory is always
exposed).

\paragraph{Fence}
\mpi{Win\_fence}, called collectively by all processes, finishes the previous exposure and access epoch and
opens the next exposure and access epoch for the whole window. An
implementation must guarantee that all remote memory operations are
committed before it leaves the fence call. Our
implementation uses an x86 \verb~mfence~ instruction (XPMEM) and DMAPP
bulk synchronization (gsync) followed by an MPI barrier to ensure global completion.
The asymptotic memory bound is $\mathcal{O}(1)$ and, assuming a good
barrier implementation, the time bound is $\mathcal{O}(\log p)$.

\paragraph{General Active Target Synchronization}
synchronizes a subset of processes of a window. Exposure
(\mpi{Win\_post}/\mpi{Win\_wait}) and access epochs
(\mpi{Win\_start}/\mpi{Win\_complete}) can be opened and closed
independently. However, a group argument is associated with each call
that starts an epoch and it states all processes participating in the
epoch. The calls have to guarantee correct \emph{matching}, i.e., if a
process $i$ specifies a process $j$ in the group argument of the post
call, then the next start call at process $j$ that has process $i$ in
the group argument \emph{matches} the post call. 

Since our RMA implementation cannot assume buffer space for remote
messages, it has to ensure
that all processes in the group argument of the start call have called a
matching post before the start returns. Similarly, the wait call has to
ensure that all matching processes have called complete. Thus, calls to 
\mpi{Win\_start} and \mpi{Win\_wait} may block, waiting for the remote
process. Both synchronizations are required to ensure integrity of the
accessed data during the epochs. The MPI specification forbids matching
configurations where processes wait cyclically (deadlocks).

\begin{figure} [h!t]
  \centering
  \begin{subfigure}{\columnwidth}
    \centering
    \includegraphics[width=0.90\linewidth]{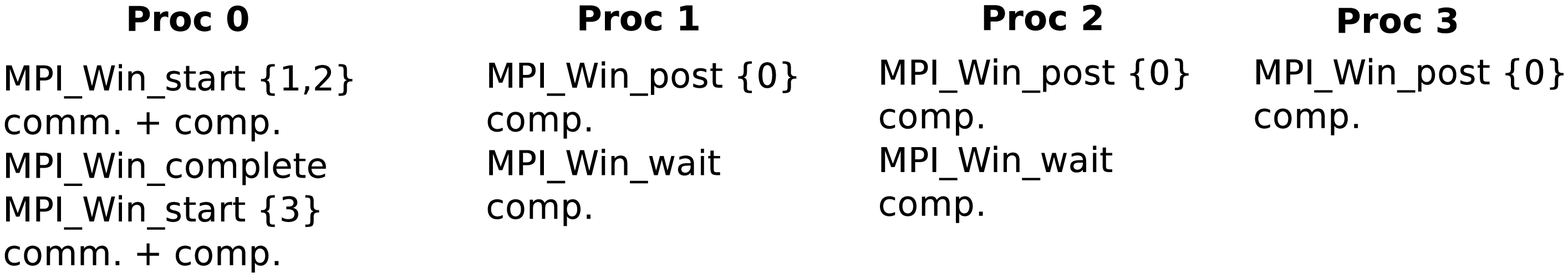}
    \caption{Source Code}
    \label{fig:pscw_example:code}
  \end{subfigure}
  \begin{subfigure}{\columnwidth}
    \centering
    \includegraphics[width=0.75\linewidth]{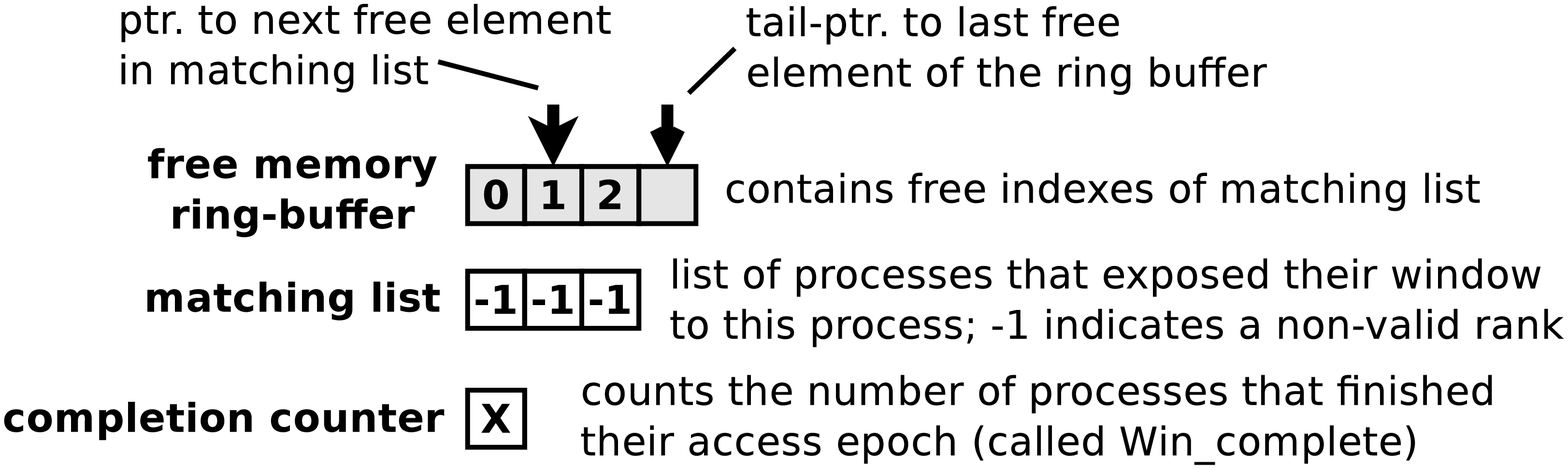}
    \caption{Data Structures}
    \label{fig:pscw_example:data}
  \end{subfigure}
  \begin{subfigure}{\columnwidth}
    \centering
    \includegraphics[width=0.50\linewidth]{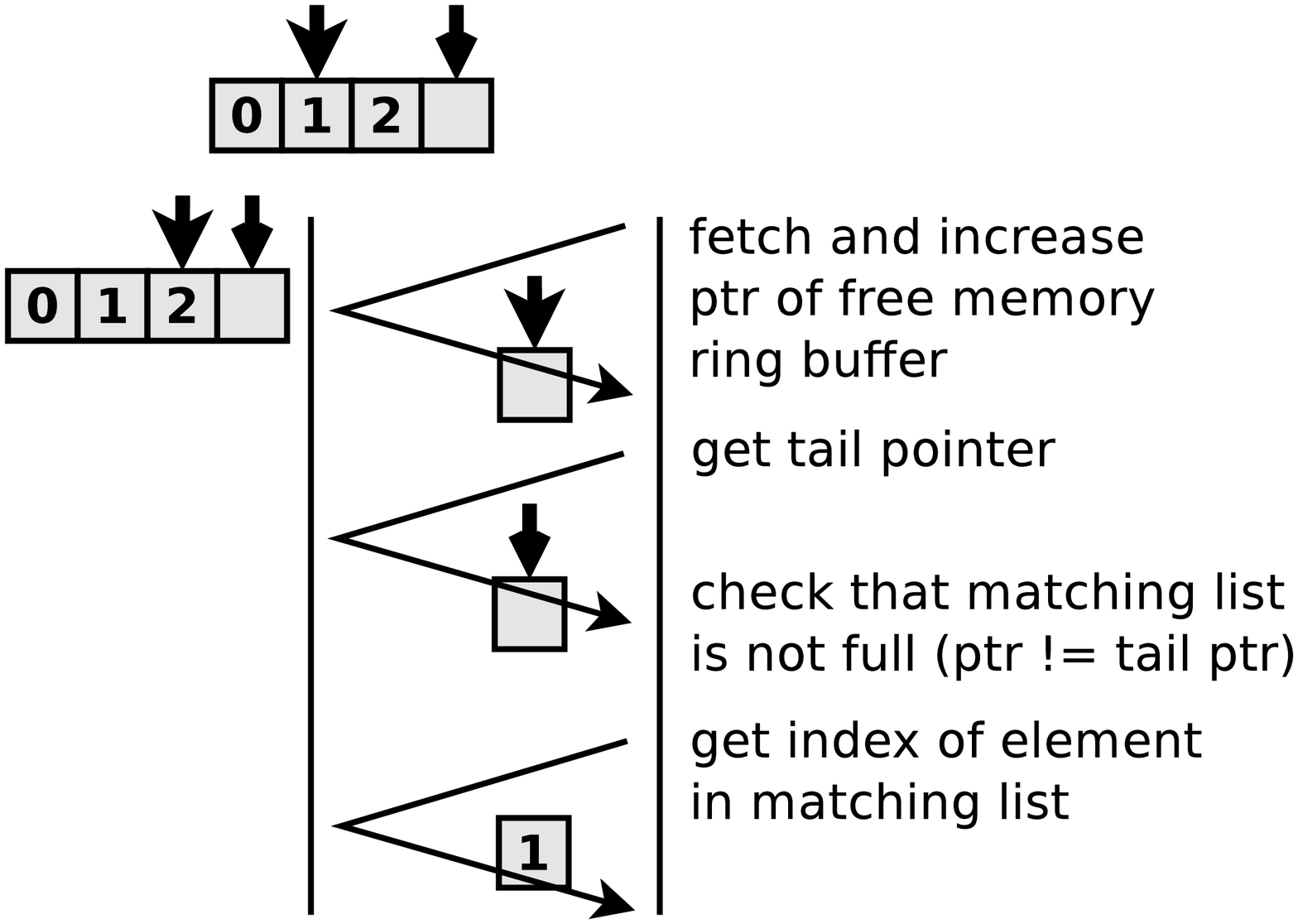}
    \caption{Free-Storage Management: Protocol to acquire a free element
    in a remote matching list, denoted as "free-mem" below}
    \label{fig:pscw_example:expl}
  \end{subfigure}
  \begin{subfigure}{\columnwidth}
    \centering
    \includegraphics[width=0.93\linewidth]{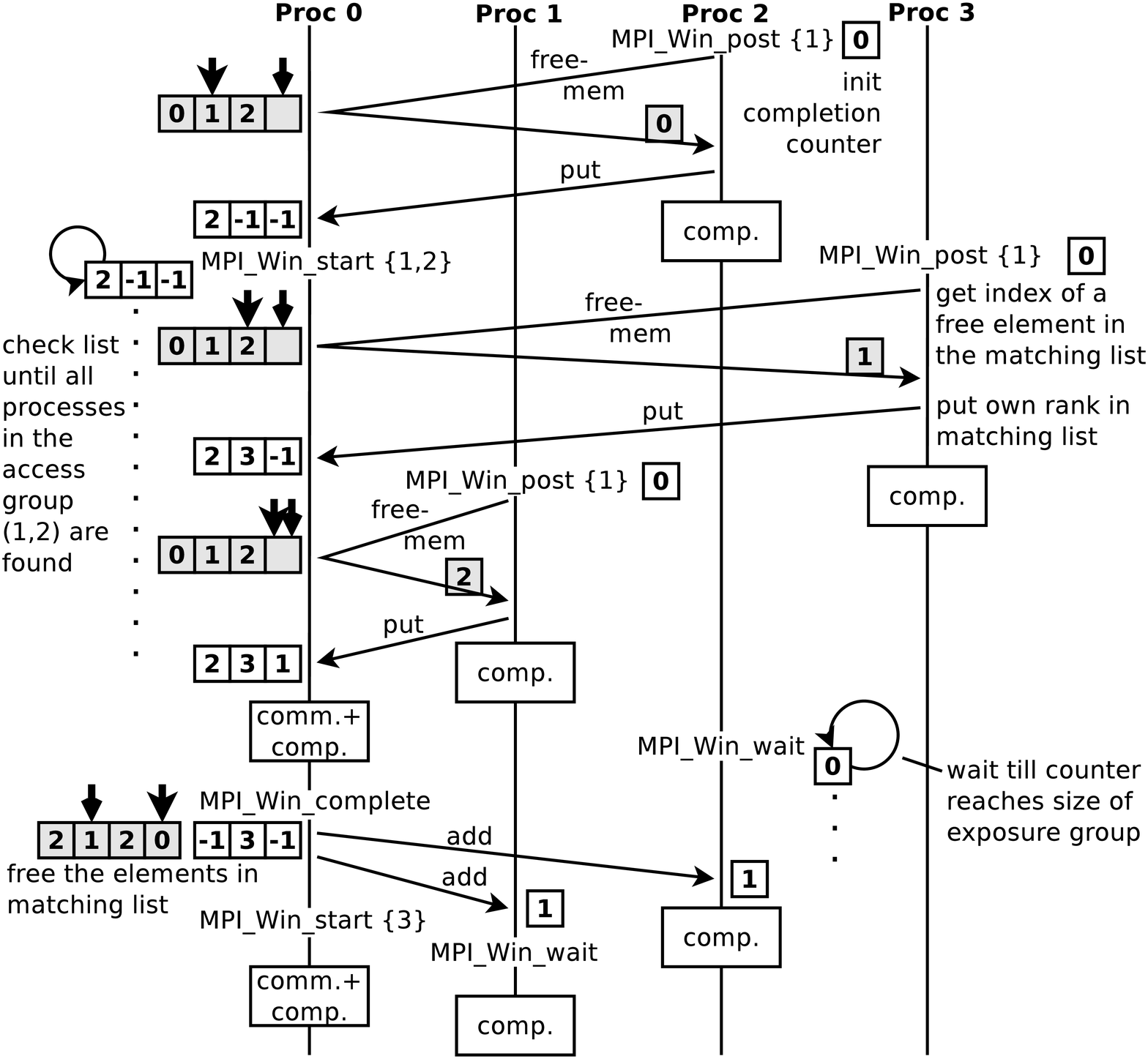}
    \caption{Possible Execution of the Complete Protocol}
    \label{fig:pscw_example:protocol}
  \end{subfigure}
\caption{Example of General Active Target Synchronization. The numbers
in the brackets for \mpi{Win\_start} and \mpi{Win\_post} indicate the
processes in the access or exposure group.}
\label{fig:pscw_example}
\end{figure}

We now describe a scalable implementation of the matching protocol with
a time and memory complexity of
$\mathcal{O}(k)$ if each process has at most $k$ neighbors across all
epochs. In addition, we assume $k$ is known to the implementation.
The scalable algorithm can be described at a high level as follows: each process $i$ that
\emph{posts} an epoch announces itself to all processes $j_1,\ldots,j_l$ in
the group argument by adding $i$ to a list local to the processes
$j_1,\ldots,j_l$. Each process $j$ that tries to \emph{start} an epoch waits
until all processes $i_1,\ldots,i_m$ in the group argument are present
in its local list. The main complexity lies in the scalable storage of this
neighbor list, needed for \emph{start}, which requires a remote free-storage
management scheme (see Figure~\ref{fig:pscw_example:expl}).
The \emph{wait} call can simply be synchronized with a 
completion counter. A process calling \emph{wait} will not return 
until the completion counter reaches the number of processes in the
specified group. To enable this, the \emph{complete} call first guarantees
remote visibility of all issued RMA operations (by calling \verb~mfence~
or DMAPP's gsync) and then increases the completion counter at
all processes of the specified group. 

Figure~\ref{fig:pscw_example:code} shows an example program with two
distinct matches to access three processes from process 0. The first
epoch on process 0 matches with processes 1 and 2 and the second epoch
matches only with process 3. Figure~\ref{fig:pscw_example:data} shows
the necessary data structures, the free memory buffer, the matching
list, and the completion counter. Figure~\ref{fig:pscw_example:expl}
shows the part of the protocol to acquire a free element in a remote
matching list and Figure~\ref{fig:pscw_example:protocol} shows a
possible execution of the complete protocol on four processes.

If $k$ is the size of the group, then the number of messages issued by
post and complete is $\mathcal{O}(k)$ and zero for start and wait. We
assume that $k\in \mathcal{O}(\log p)$ in scalable
programs~\cite{242438}.

\paragraph{Lock Synchronization} 

We now describe a low-overhead and scalable strategy to implement shared global, and
shared and exclusive process-local locks on RMA systems (the MPI-3.0
specification does not allow exclusive global lock all). We utilize a two-level
lock hierarchy: 
one global lock variable (at a designated process, called \emph{master})
and $p$ local lock variables (one lock on each process). We assume that
the word-size of the machine, and thus each lock variable, is 64 bits.
Our scheme also generalizes to other $t$ bit word sizes as long as the
number of processes is not more than $2^{\left\lfloor\frac{t}{2}\right\rfloor}$.

\begin{figure}[h!t] 
  \centering
  \begin{subfigure}{\columnwidth}
    \centering
    \includegraphics[width=0.75\linewidth]{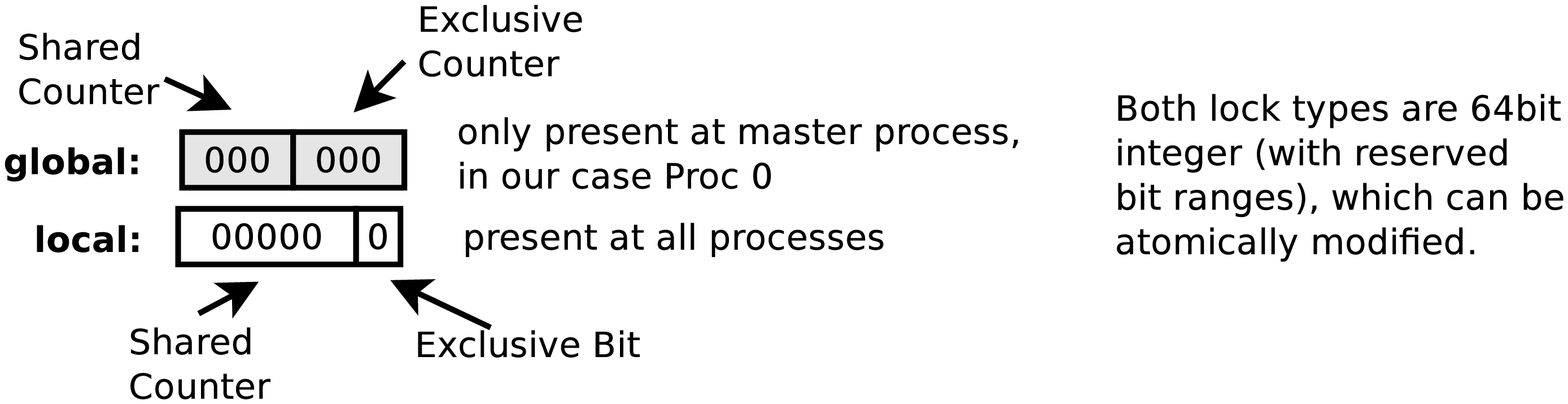}
    \caption{Data Structures}
    \label{fig:locks:data}
  \end{subfigure}
  \begin{subfigure}{\columnwidth}
    \centering
    \includegraphics[width=0.75\linewidth]{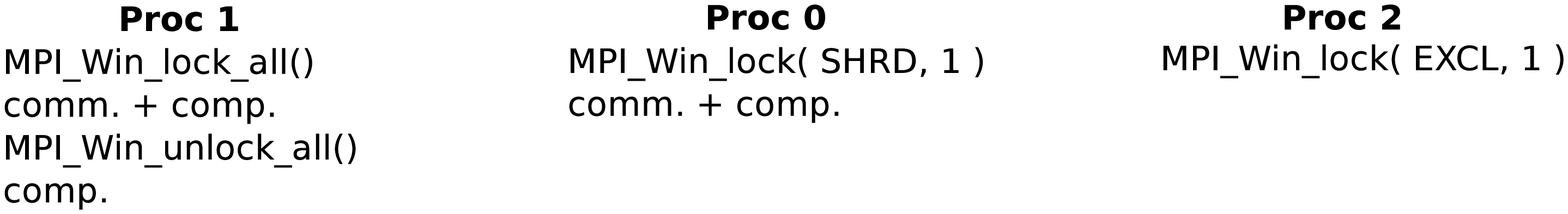}
    \caption{Source Code}
    \label{fig:locks:code}
  \end{subfigure}
  \begin{subfigure}{\columnwidth}
    \centering
    \includegraphics[width=0.90\linewidth]{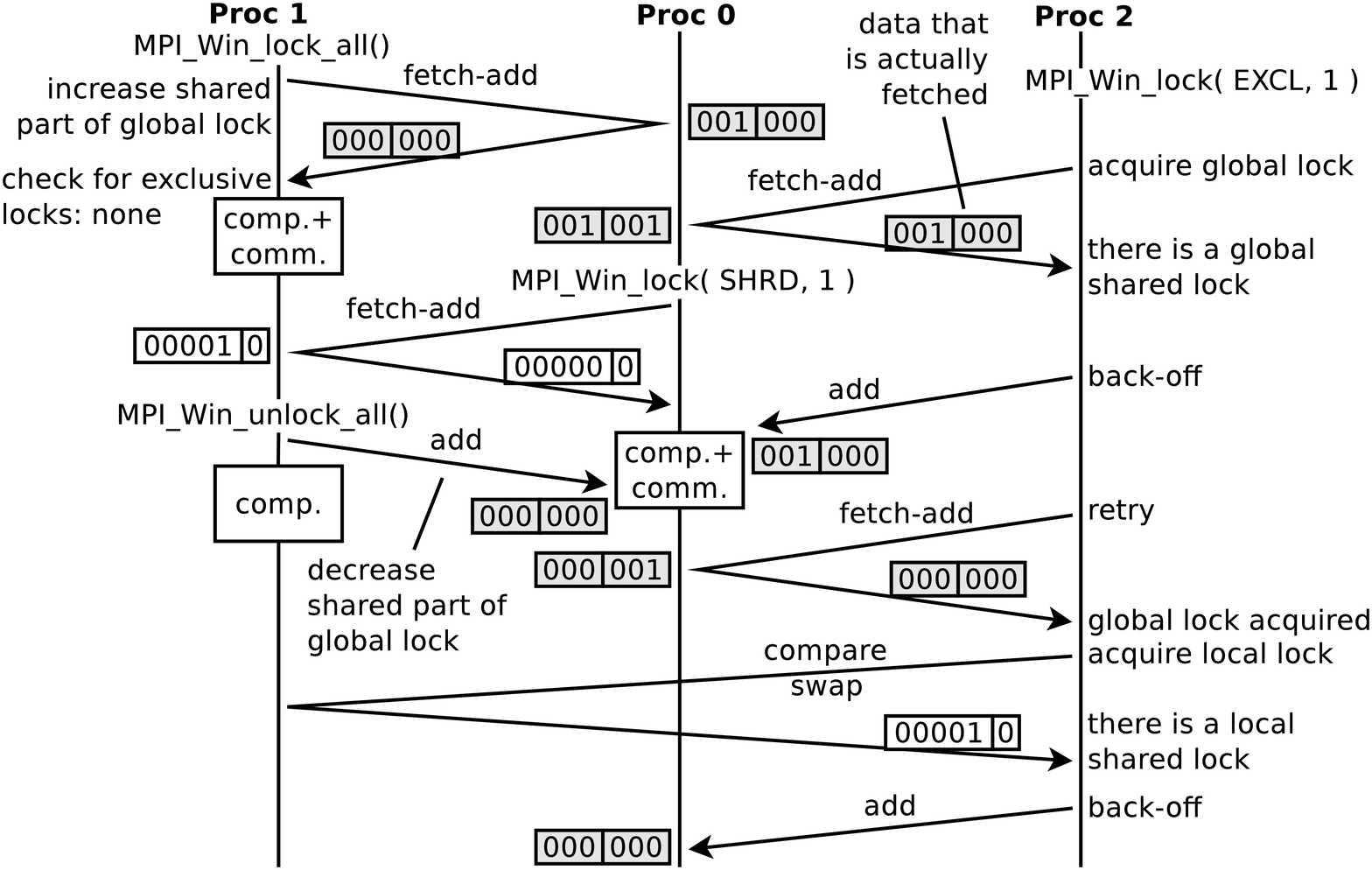}
    \caption{Possible Schedule}
    \label{fig:locks:protocol}
  \end{subfigure}
  \caption{Example of Lock Synchronization} 
  \label{fig:locks} 
\end{figure}

Each local lock variable is used to implement a reader-writer lock,
which allows only one writer (\emph{exclusive} lock), but many readers
(\emph{shared} locks). The highest order bit of the lock variable
indicates a write access, while the other bits are used to count the
number of shared locks held by other processes
(cf.~\cite{Mellor-Crummey:1991:SRS:109626.109637}).
The global lock variable is split into two parts. The first part counts
the number of processes holding a global shared lock in the window and
the second part counts the number of exclusively locked processes. Both
parts guarantee that there are only accesses of one type (either
exclusive or lock-all) concurrently. This data structure enables all
lock operations to complete in $\mathcal{O}(1)$ steps if a lock can be
acquired immediately.
Figure~\ref{fig:locks:data} shows the structure of the local and global
lock variables (counters).

Figure~\ref{fig:locks:code} shows an exemplary lock scenario for three
processes. 
We do not provide an algorithmic description of the protocol due to the
lack of space (the source-code is available online). However, we
describe a locking scenario to foster understanding of the protocol.
Figure~\ref{fig:locks:protocol} shows a possible execution schedule for
the scenario from Figure~\ref{fig:locks:code}. Please note that we
permuted the order of processes to (1,0,2) instead of the intuitive 
(0,1,2) to minimize overlapping lines in the figure.

Process 1 starts a lock all epoch by increasing the global shared
counter atomically.  Process 2 has to wait with its exclusive lock
request until Process 1 finishes its lock all epoch. The waiting is
performed by an atomic fetch and add. If the result of this operation
finds any global shared lock then it backs off its request and does not
enter the lock, and proceeds to retry. If there is no global shared lock
then it enters its local lock phase.
An acquisition of a shared lock on a specific target (\mpi{Win\_lock})
only involves the local lock on the target. The origin process (e.g.,
Process 0) fetches and increases the lock in one atomic operation. If
the writer bit is not set, the origin can proceed.  If an exclusive lock
is present, the origin repeatedly (remotely) reads the lock until the
writer finishes his access. All waits/retries can be performed with
exponential back off to avoid congestion.

Summarizing the protocol: For a local exclusive lock, the origin process
needs to ensure two invariants: (1) no global shared lock can be held or
acquired during the local exclusive lock {\bf and} (2) no local shared
or exclusive lock can be held or acquired during the local exclusive
lock. For the first part, the locking process fetches the lock variable
from the master process and also increases the writer part in one atomic
operation to register its wish for an exclusive lock. If the fetched
value indicates lock all accesses, then the origin backs off by
decreasing the writer part of the global lock. In case there is no
global reader, the origin proceeds to the second invariant and tries to
acquire an exclusive local lock on its target using compare-and-swap
with zero (cf.~\cite{Mellor-Crummey:1991:SRS:109626.109637}). If this
succeeds, the origin acquired the lock and can proceed. In the example,
Process 2 succeeds at the second attempt to acquire the global lock but
fails to acquire the local lock and needs to back off by releasing its
exclusive global lock. Process 2 will repeat this two-step operation
until it acquired the exclusive lock. If a process already holds any
exclusive lock, then it can immediately proceed to invariant two.

When unlocking (\mpi{Win\_unlock}) a shared lock, the origin only has to
atomically decrease the local lock on the target. In case of an
exclusive lock it requires two steps. The first step is the same as in
the shared case, but if the origin does not hold any additional
exclusive locks, it has to release its global lock by atomically
decreasing the writer part of the global lock.

The acquisition or release of a shared lock on all processes of the
window (\mpi{Win\_lock\_all}/\mpi{Win\_unlock\_all}) is similar to the
shared case for a specific target, except it targets the global 
lock.

If no exclusive locks exist, then shared locks (both \mpi{Win\_lock} and
\mpi{Win\_lock\_all}) only take one remote atomic update operation.
The number of remote requests while waiting can be bound by using MCS
locks~\cite{Mellor-Crummey:1991:SWC:106973.106999}. The first exclusive
lock will take in the best case two atomic communication operations.
This will be reduced to one atomic operation if the origin process
already holds an exclusive lock. Unlock operations always cost one
atomic operation, except for the last unlock in an exclusive case with
one extra atomic operation for releasing the global lock. The memory
overhead for all functions is $\mathcal{O}(1)$.

\paragraph{Flush} Flush guarantees remote completion and
is thus one of the most performance-critical functions on MPI-3.0 RMA
programming. \fompi{}'s flush implementation relies on the underlying
interfaces and simply issues a DMAPP remote bulk completion and an x86
\verb~mfence~. All flush operations (\mpi{Win\_flush},
\mpi{Win\_flush\_local}, \mpi{Win\_flush\_all}, and
\mpi{Win\_flush\_all\_local}) share the same implementation and add only
78 CPU instructions (x86) to the critical path. 

\subsection{Communication Functions}

Communication functions map nearly directly to low-level hardware
functions. This is a major strength of RMA programming. In \fompi{},
put and get simply use DMAPP put and get for remote
accesses or local memcpy for XPMEM accesses. Accumulates either use
DMAPP atomic operations (for many common integer operations on 8 Byte
data) or fall back to a simple protocol that locks the remote
window, gets the data, accumulates it locally, and writes it back.
This fallback protocol is necessary to avoid involvement of the
receiver for true passive mode. It can be improved if we allow
buffering (enabling a space-time
trade-off~\cite{Zhao:2012:ASO:2404033.2404043}) such that active-mode
communications can employ active messages to perform the remote
operations atomically. 

\paragraph{Handling Datatypes}
Our implementation supports arbitrary MPI datatypes by using the
MPITypes library~\cite{mpitypes}. In each communication, the datatypes
are split into the smallest number of contiguous blocks (using both, the
origin \emph{and} target datatype) and one DMAPP operation or memory
copy (XPMEM) is initiated for each block. 

While offering the full functionality of the rich MPI interface, our
implementation is highly tuned for the common case of contiguous data
transfers using intrinsic datatypes (e.g.,~\mpi{DOUBLE}). Our full 
implementation adds only 173 CPU instructions (x86) in the optimized
critical path of \mpi{Put} and \mpi{Get}. We also utilize SSE-optimized
assembly code to perform fast memory copies for XPMEM communication.

\begin{figure*}[h!t]
\centering
\begin{subfigure}{0.32\textwidth}
  \includegraphics[width=\columnwidth]{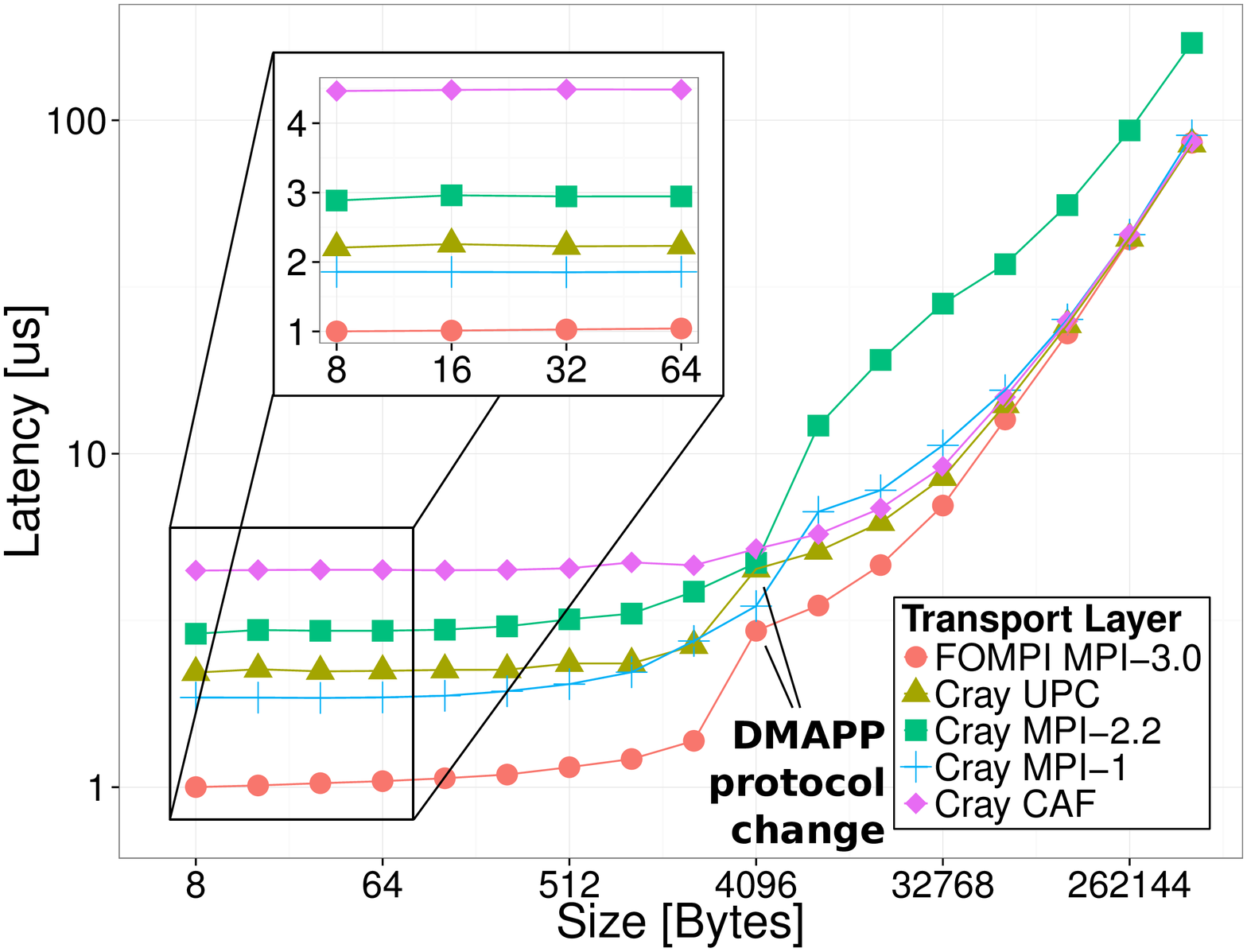}
  \caption{Latency inter-node Put}
  \label{fig:microbench:latency:put:offnode}
\end{subfigure}
\begin{subfigure}{0.32\textwidth}
  \includegraphics[width=\columnwidth]{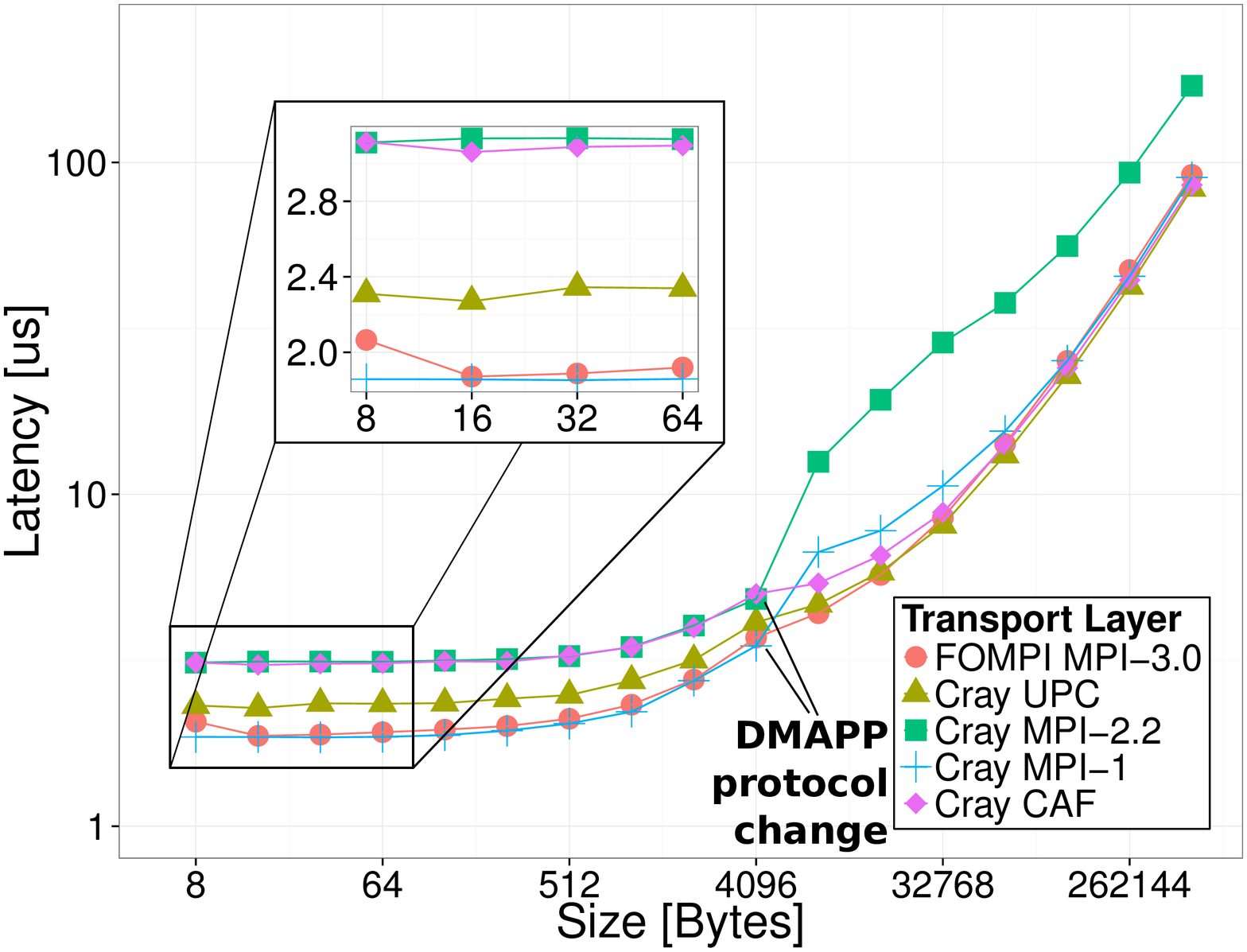}
  \caption{Latency inter-node Get}
  \label{fig:microbench:latency:get:offnode}
\end{subfigure}
\begin{subfigure}{0.32\textwidth}
  \includegraphics[width=\columnwidth]{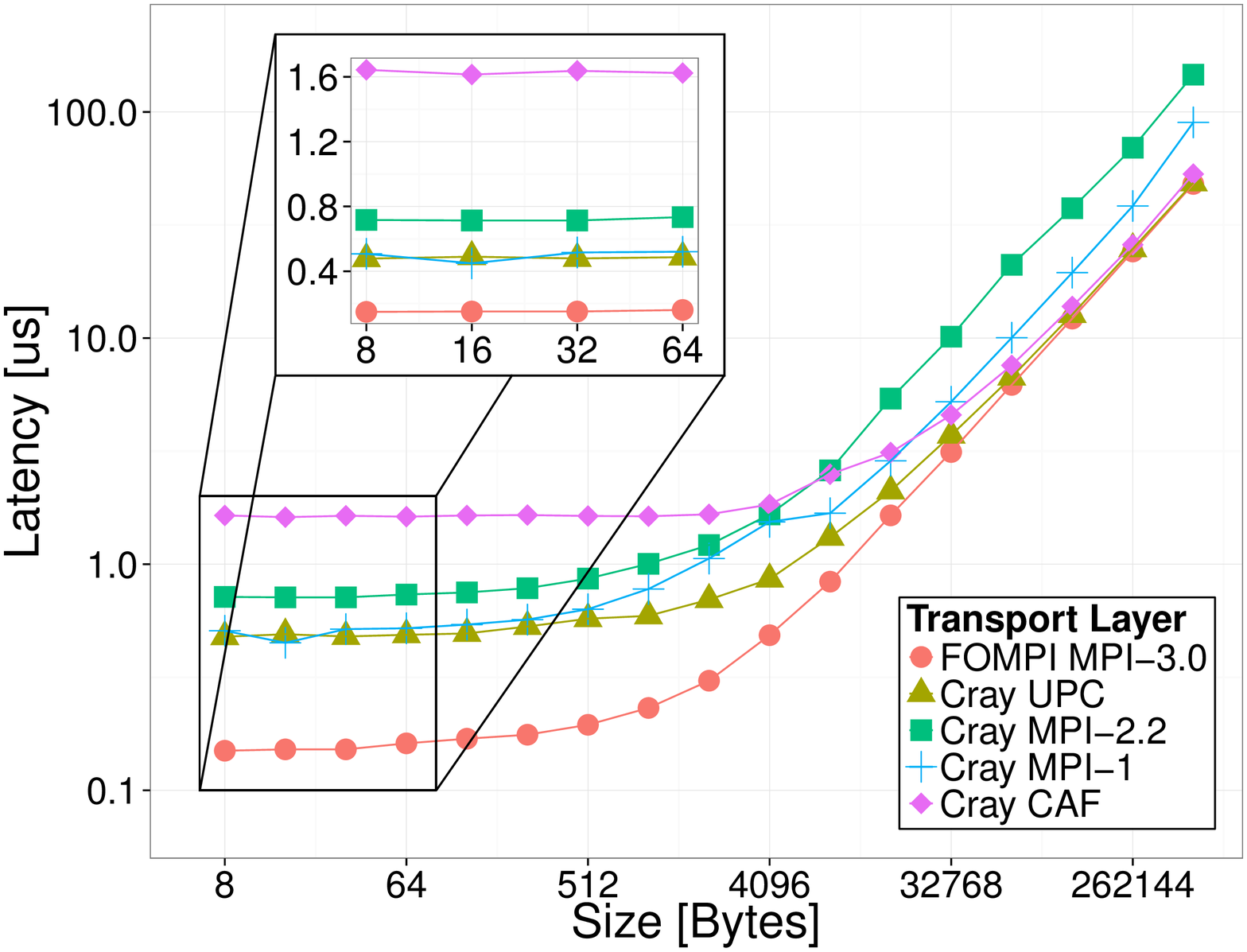}
  \caption{Latency intra-node Put/Get}
  \label{fig:microbench:latency:onnode}
\end{subfigure}
\caption{Latency comparison for remote put/get for DMAPP and XPMEM
(shared memory) communication. Note that MPI-1 Send/Recv implies
remote synchronization while UPC, Fortran Coarrays and MPI-2.2/3.0 only guarantee
consistency.}
\label{fig:microbench:latency}
\end{figure*}

\begin{figure*}[h!t]
\centering
\begin{subfigure}{0.32\textwidth}
  \includegraphics[width=\columnwidth]{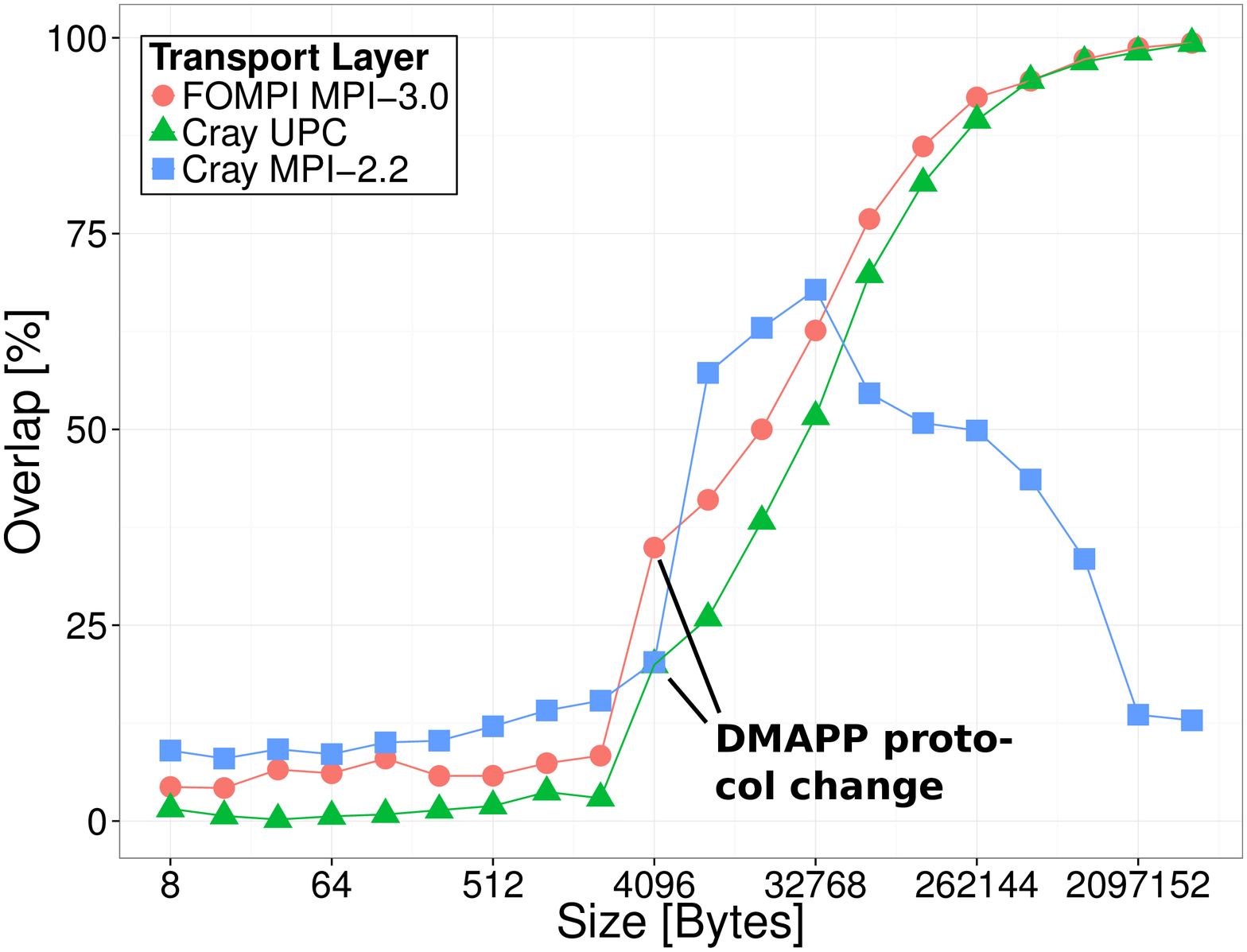}
  \caption{Overlap inter-node}
  \label{fig:microbench:overlap}
\end{subfigure}
\begin{subfigure}{0.32\textwidth}
  \includegraphics[width=\columnwidth]{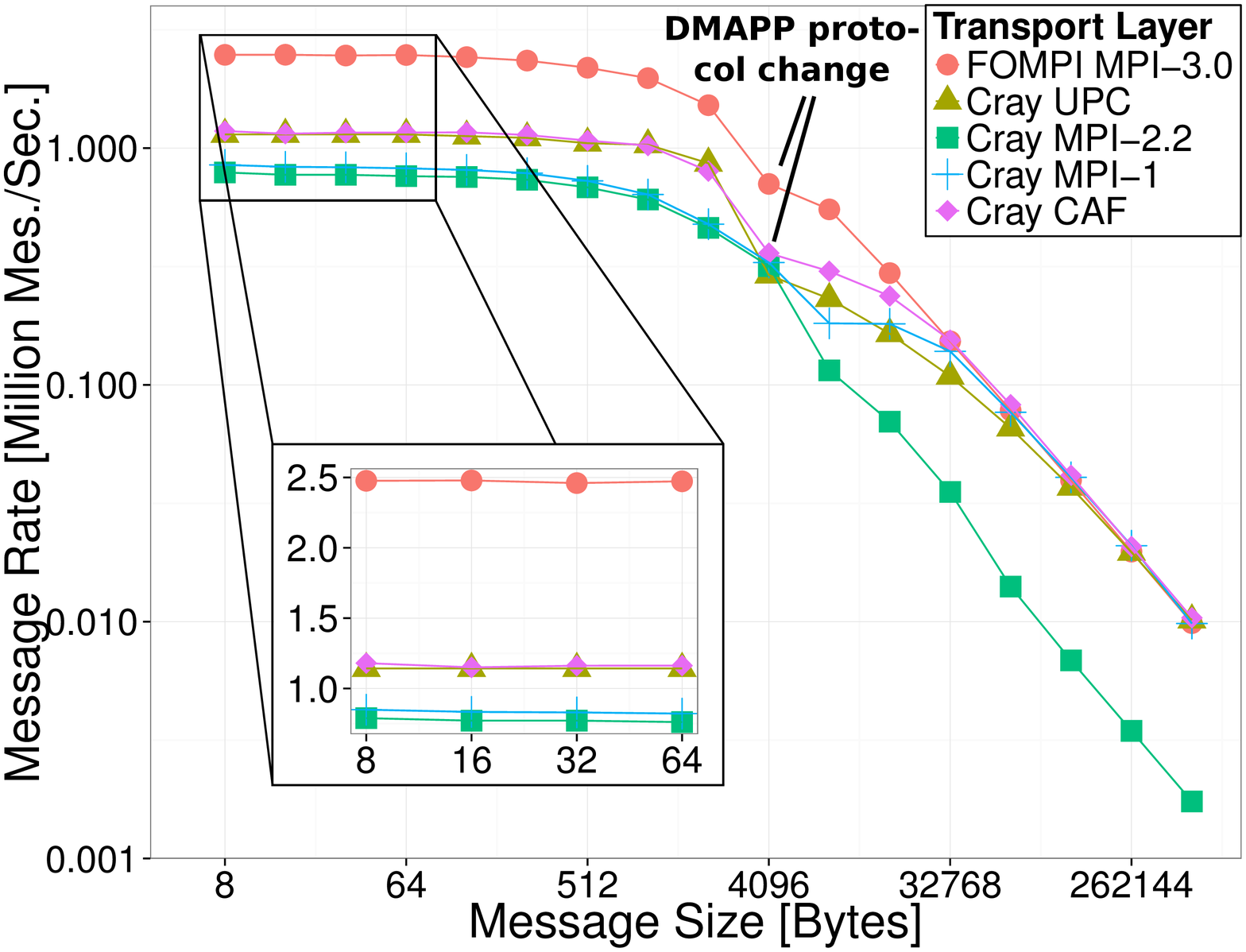}
  \caption{Message Rate inter-node}
  \label{fig:microbench:message_rate:offnode}
\end{subfigure}
\begin{subfigure}{0.32\textwidth}
  \includegraphics[width=\columnwidth]{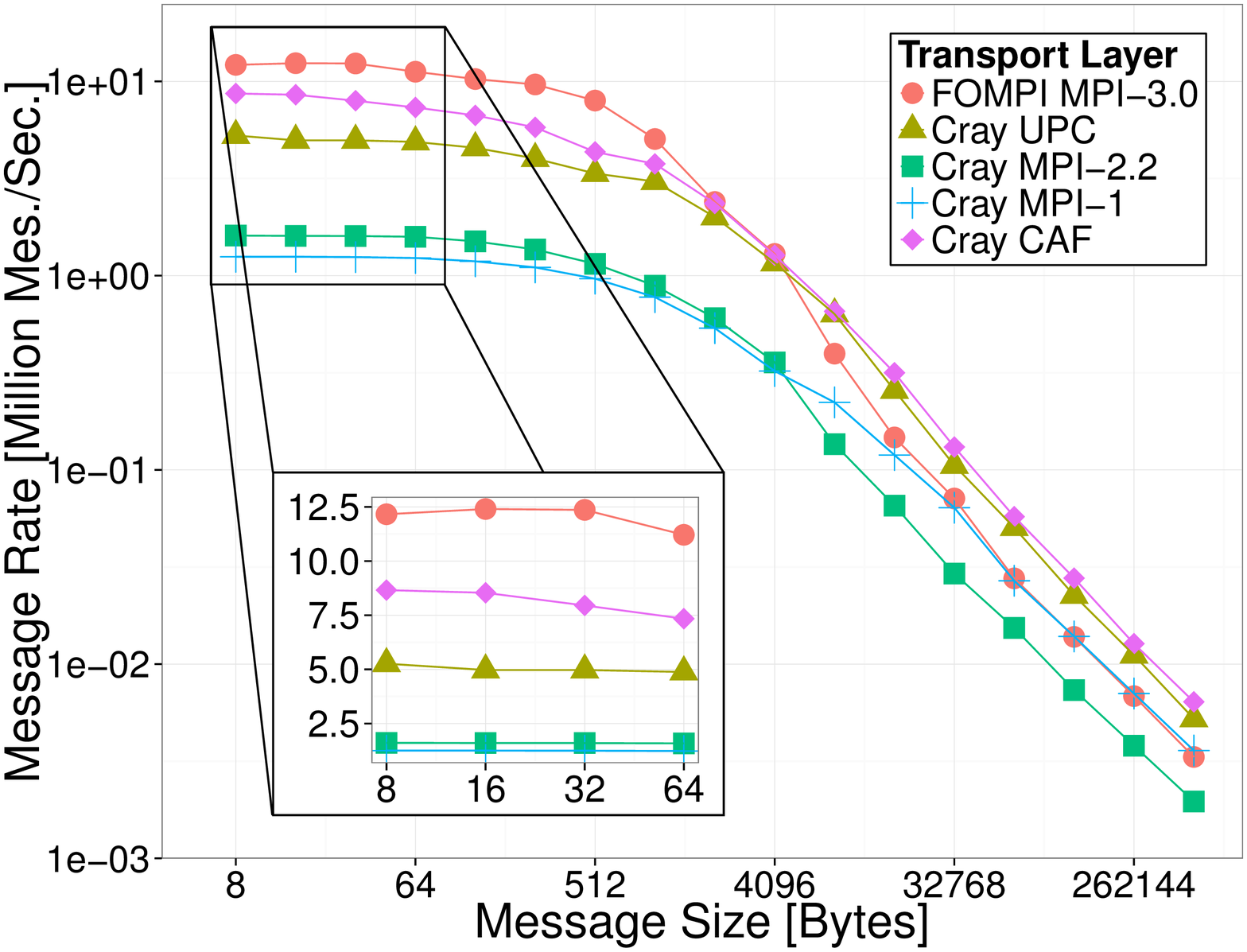}
  \caption{Message Rate intra-node}
  \label{fig:microbench:message_rate:onnode}
\end{subfigure}
\caption{Communication/computation overlap for Put over DMAPP, Cray MPI-2.2
  has much higher latency up to 64 kB (cf.
  Figure~\ref{fig:microbench:latency:put:offnode}), thus allows higher
  overlap. XPMEM
implementations do not support overlap due to the shared memory copies.
Figures (b) and (c) show message rates for put communication for all
transports.}
\label{fig:microbench:message_rate}
\end{figure*}

\subsection{Blocking Calls}

The MPI standard allows an implementation to block in several
synchronization calls. Each correct MPI program should thus never
deadlock if all those calls are blocking. However, if the user knows
the detailed behavior, she can tune for performance, e.g., if locks
block, then the user may want to keep lock/unlock regions short.
We describe here which calls may block depending on other processes and
which calls will wait for other processes to reach a certain state.
We point out that, in order to write (performance)
portable programs, the user cannot rely on such knowledge in general!

With our protocols, (a) \mpi{Win\_fence} waits for all other window processes to
enter the \mpi{Win\_fence} call, (b) \mpi{Win\_start} waits for matching
\mpi{Win\_post} calls from all processes in the access group, (c)
\mpi{Win\_wait} waits for the calls to \mpi{Win\_complete} from all
processes in the exposure group, and (d) \mpi{Win\_lock} and
\mpi{Win\_lock\_all} wait until they acquired the desired lock.

\section{Detailed Performance Modeling and Evaluation}
\label{sec:micro}

We now describe several performance features of our protocols and
implementation and compare it to Cray MPI's highly tuned point-to-point
as well as its relatively untuned one sided communication. In addition,
we compare \fompi{} with two major HPC PGAS languages: UPC and Fortran
2008 with Coarrays, both specially tuned for Cray systems. 
We did not evaluate the semantically richer Coarray Fortran
2.0~\cite{Mellor-Crummey:2009:NVC:1809961.1809969} because no tuned
version was available on our system.
We execute all benchmarks on the Blue Waters system, using the Cray XE6
nodes only. Each compute node contains four 8-core AMD Opteron 6276
(Interlagos) 2.3~GHz and is connected to other nodes through a 3D-Torus
Gemini network. We use the Cray Programming Environment 4.1.40 for MPI,
UPC, and Fortran Coarrays, and GNU gcc 4.7.2 when features that are not
supported by Cray's C compiler are required (e.g., inline assembly for a
fast x86 SSE copy loop). 

Our benchmark measures the time to perform a single operation (for a
given process count and/or data size) at all processes and adds the
maximum across all ranks to a bucket. The run is repeated 1,000
times to gather statistics. We use the cycle accurate x86 \verb~RDTSC~
counter for each measurement. 
All performance figures show the medians of all gathered points for each
configuration. 

\subsection{Latency and Bandwidth}

Comparing latency and bandwidth between one sided RMA communication and
point-to-point communication is not always fair since RMA communication
may require extra synchronization to notify the target. All latency
results presented for RMA interfaces are guaranteeing remote completion
(the message is committed in remote memory) but no synchronization. We
analyze synchronization costs separately in Section~\ref{sec:sync}. 

We measure MPI-1 point-to-point latency with standard ping-pong
techniques. 
For Fortran Coarrays, we use a remote assignment of a double precision array
of size \verb~SZ~:
\begin{lstlisting}[language=Fortran,basicstyle=\small,frame=tb,keywordstyle=\color{black}\bfseries,morekeywords={sync,memory}]
double precision, dimension(SZ) :: buf[*]
do memsize in all sizes<SZ
  buf(1:memsize)[2] = buf(1:memsize)
  sync memory
end do
\end{lstlisting}

In UPC, we use a single shared array and the intrinsic function memput,
we also tested shared pointers with very similar performance:
\begin{lstlisting}[language=C,basicstyle=\small,frame=tb,keywordstyle=\color{black}\bfseries,morekeywords={shared,upc_fence,upc_memput,upc_all_alloc}]
shared [SZ] double *buf;
buf = upc_all_alloc(2, SZ);
for(size=1 ; size<=SZ ; size*=2) {
  upc_memput(&buf[SZ], &priv_buf[0], size);
  upc_fence;
}
\end{lstlisting}

In MPI-3.0 RMA, we use an allocated window and passive target mode with
flushes:
\begin{lstlisting}[language=C,basicstyle=\small,frame=tb,keywordstyle=\color{black}\bfseries,morekeywords={MPI_Win_allocate,MPI_Win_lock,MPI_Win_unlock,MPI_Put,MPI_Win_flush}]
MPI_Win_allocate(SZ, ..., &buf, &win);
MPI_Win_lock(excl, 1, ..., win);
for(size=1 ; size<=SZ ; size*=2) {
  MPI_Put(&buf[0], size, ..., 1, ..., win);
  MPI_Win_flush(1, win);
}
MPI_Win_unlock(1, win);
\end{lstlisting}

Figures~\ref{fig:microbench:latency:put:offnode},~\ref{fig:microbench:latency:get:offnode},
and~\ref{fig:microbench:latency:onnode} show the latency for
varying message sizes for intra- and inter-node put, get. Due to the
highly optimized fast-path, \fompi{} has more than 50\% lower latency
than other PGAS models while achieving the same bandwidth for larger
messages.
The performance functions (cf. Figure~\ref{fig:overview}) are: $\mathcal{P}_{put} = 0.16
ns \cdot s + 1 \mu s$ and $\mathcal{P}_{get} = 0.17 ns \cdot s + 1.9 \mu
s$.

\subsubsection{Overlapping Computation}

The overlap benchmark measures how much of the communication time can be
overlapped with computation. It calibrates a computation loop to consume
slightly more time than the latency.  Then it places computation between
the communication and the synchronization and measures the combined
time. The ratio of overlapped computation is then computed from the
measured communication, computation, and combined times.
Figure~\ref{fig:microbench:overlap} shows the ratio of the communication
that can be overlapped for Cray's MPI-2.2, UPC, and \fompi{} MPI-3.0.

\begin{figure*}[h!t]
\centering
\begin{subfigure}{0.32\textwidth}
  \includegraphics[width=\columnwidth]{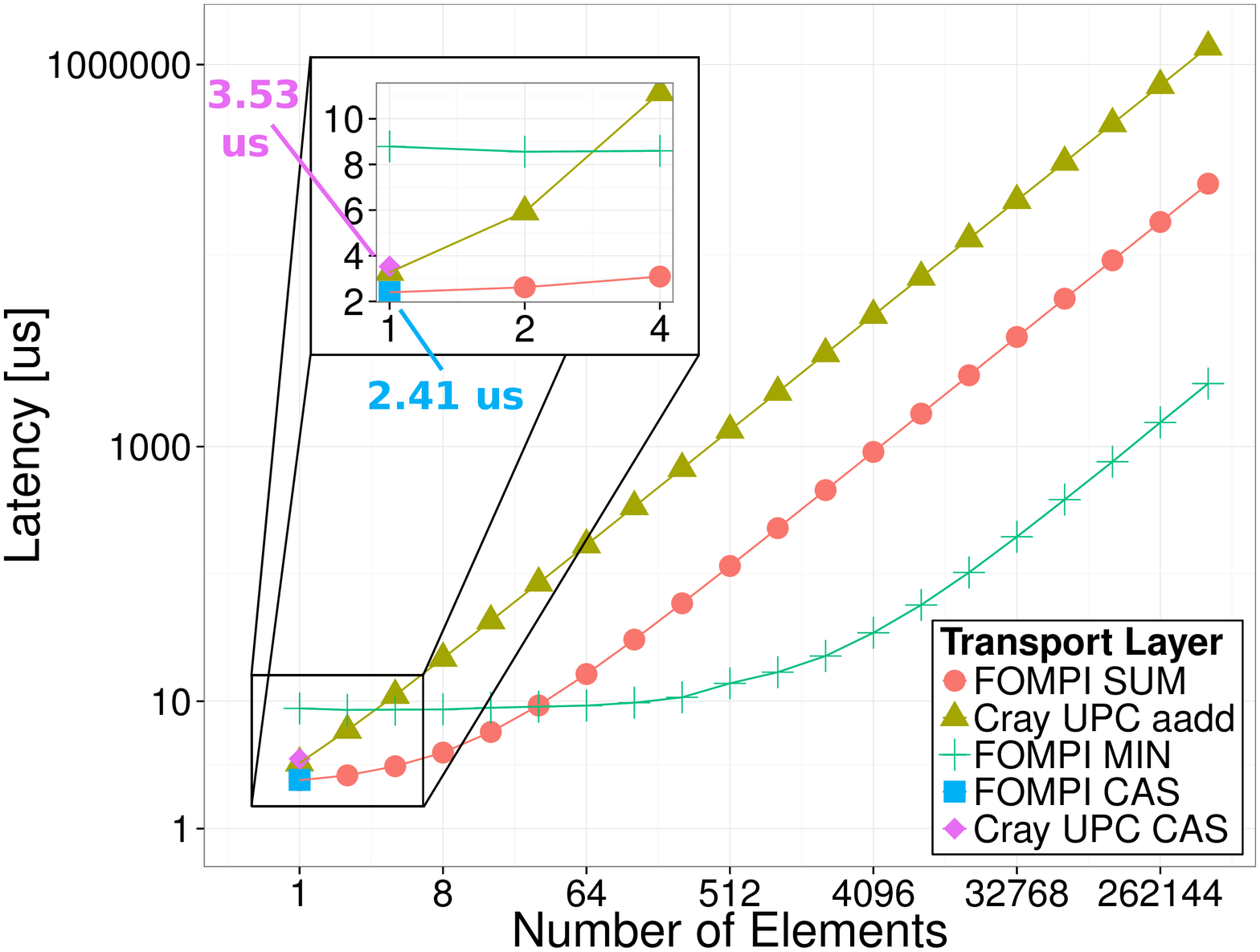}
  \caption{Atomic Operation Performance}
  \label{fig:atomics}
\end{subfigure}
\begin{subfigure}{0.31\textwidth}
  \vspace{0.2em}
  \includegraphics[width=\columnwidth]{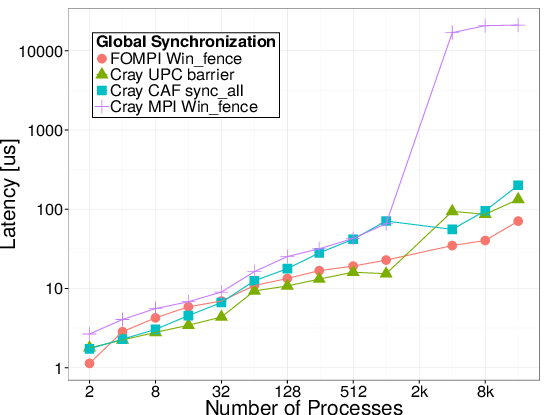}
  \caption{Latency for Global Synchronization}
  \label{fig:microbench:scale:fence}
\end{subfigure}
\begin{subfigure}{0.32\textwidth}
  \includegraphics[width=\columnwidth]{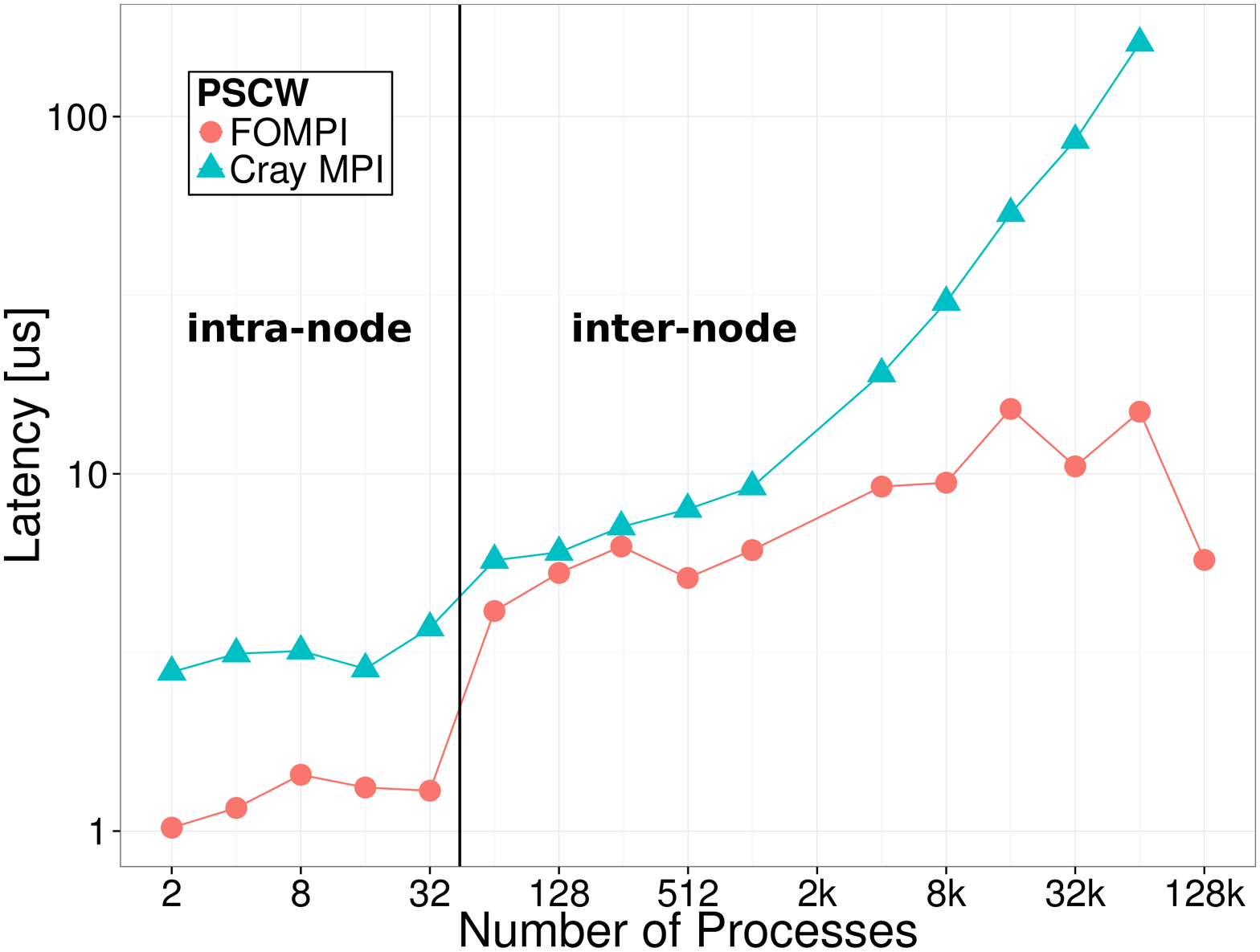}
  \caption{Latency for PSCW (Ring Topology)}
  \label{fig:microbench:scale:pscw}
\end{subfigure}
\caption{Performance of atomic accumulate
operations and synchronization latencies.}
\label{fig:microbench:scale}
\end{figure*}

\subsubsection{Message Rate}

The message rate benchmark is very similar to the latency benchmark,
however, it benchmarks the start of 1,000 transactions without
synchronization. This determines the overhead for starting a single
operation. The Cray--specific PGAS pragma \verb+defer_sync+ was used in
the UPC and Fortran Coarrays versions for full optimization.
Figures~\ref{fig:microbench:message_rate:offnode}
and~\ref{fig:microbench:message_rate:onnode} show the message rates for
DMAPP and XPMEM (shared memory) communications, respectively. Injecting
a single 8-Byte message costs $416 ns$ for inter-node and $80ns$
($\approx$190 instructions) for intra-node case.

\subsubsection{Atomics}

Figure~\ref{fig:atomics} shows the performance of the DMAPP-accelerated
\mpi{SUM} of 8-Byte elements, a non-accelerated \mpi{MIN}, and 8-Byte CAS. 
The performance functions are $\mathcal{P}_{acc,sum} = 28 ns \cdot s +
2.4 \mu s$, $\mathcal{P}_{acc,min} = 0.8 ns \cdot s +
7.3 \mu s$, and $\mathcal{P}_{CAS} = 2.4 \mu s$. The DMAPP acceleration
lowers the latency for small messages while the locked
implementation exhibits a higher bandwidth. However, this does not
consider the serialization due to the locking.

\subsection{Synchronization Schemes}
\label{sec:sync}

In this section, we study the overheads of the various synchronization
modes. The different modes have nontrivial trade-offs. For example
General Active Target Synchronization performs better if small groups of processes are
synchronized and fence synchronization performs best if the
synchronization groups are essentially as big as the full group attached
to the window. However, the exact crossover point is a function of the
implementation and system. While the active target mode notifies the target
implicitly that its memory is consistent, in passive target mode, the
user has to do this explicitly or rely on synchronization side effects
of other functions (e.g., allreduce). 

We thus study and model the performance of all synchronization modes and
notification mechanisms for passive target. Our performance models
can be used by the programmer to select the best option for the problem
at hand. 

\paragraph{Global Synchronization}

Global synchronization is offered by fences in MPI-2.2 and MPI-3.0. It can
be directly compared to Fortran Coarrays \verb~sync all~ and UPC's
\verb~upc_barrier~ which also synchronize the memory at all processes.
Figure~\ref{fig:microbench:scale:fence}
compares the performance of \fompi{} with Cray's MPI-2.2, UPC, and
Fortran Coarrays implementations.
The performance function for \fompi{}'s fence
implementation is:
$\mathcal{P}_{fence} =  2.9\mu s \cdot \log_2(p)$.

\begin{figure*}[h!t]
\centering
\begin{subfigure}{0.32\textwidth}
\includegraphics[width=\columnwidth]{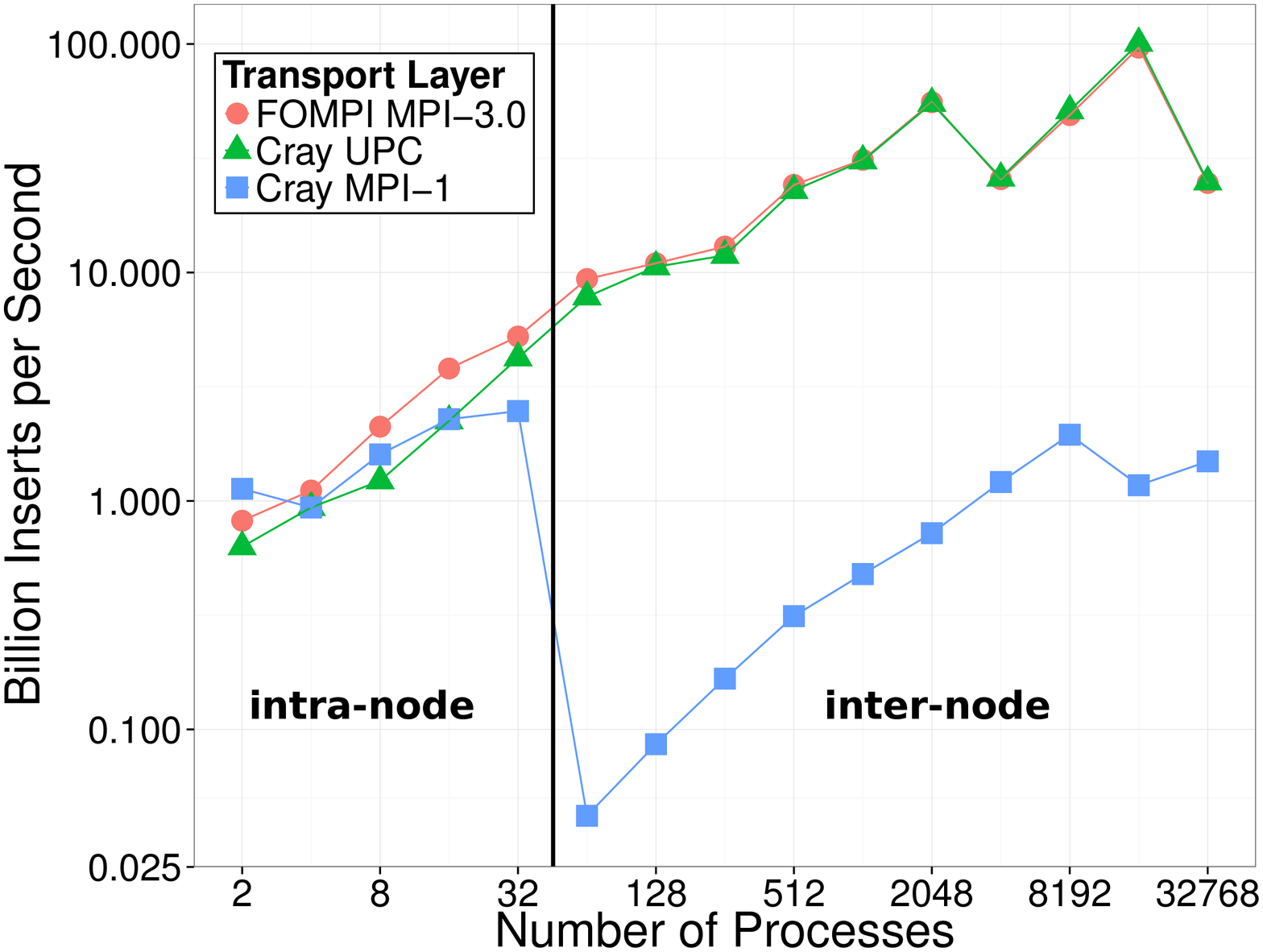}
  \caption{Inserts per second for inserting 16k elements per process including synchronization.}
  \label{fig:hashtablePlots}
\end{subfigure}
\begin{subfigure}{0.31\textwidth}
  \includegraphics[width=\columnwidth]{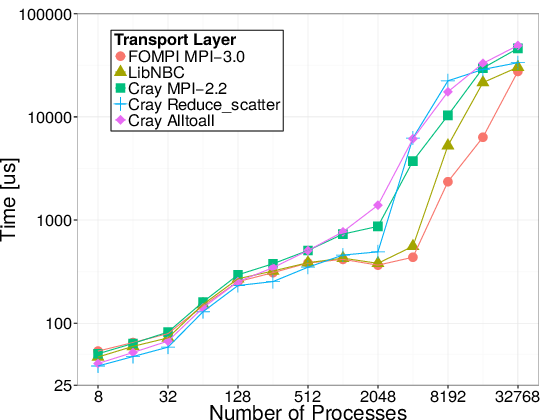}
  \caption{Time to perform one dynamic sparse data exchange (DSDE) with 6
  random neighbors}
  \label{fig:applications:dsde}
\end{subfigure}
\begin{subfigure}{0.32\textwidth}
   \includegraphics[width=\columnwidth]{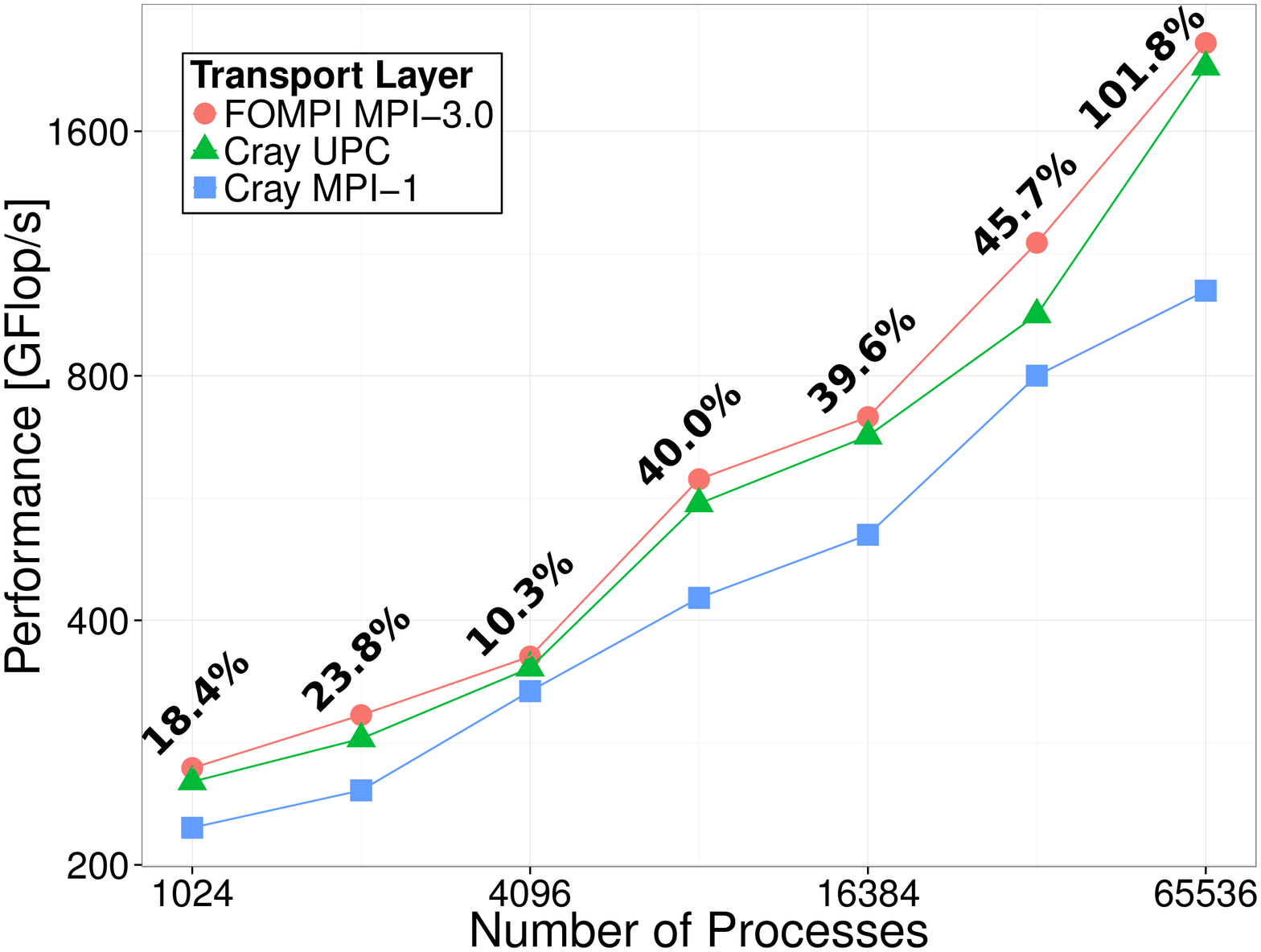}
   \caption{3D FFT Performance. The annotations represent the
      improvement of \fompi{} over MPI-1.}
   \label{fig:applications:fft}
\end{subfigure}
\caption{Application Motifs: (a) hashtable representing data
analytics applications and key-value stores, (b) dynamic sparse data
exchange representing graph traversals, n-body methods, and rapidly
evolving meshes~\cite{hoefler-dsde}, (c) full parallel 3D FFT}
\label{fig:applications:results}
\end{figure*}

\paragraph{General Active Target Synchronization}

Only MPI-2.2 and MPI-3.0 offer General Active Target (also called ``PSCW'')
synchronization. 
A similar mechanism (\verb+sync images+) for Fortran Coarrays was
unfortunately not available on our test system.
Figure~\ref{fig:microbench:scale:pscw} shows the performance for Cray
MPI and \fompi{} when synchronizing an one-dimensional Torus (ring) where
each process has exactly two neighbors (k=2). An ideal implementation
would exhibit constant time for this benchmark. We observe
systematically growing overheads in Cray's implementation as well as
system
noise~\cite{Petrini:2003:CMS:1048935.1050204,Hoefler:2010:CIS:1884643.1884668}
on runs with more than 1,000 processes with \fompi{}.
We model the performance with varying numbers of neighbors and
\fompi{}'s PSCW synchronization costs involving $k$ off-node neighbor are
$\mathcal{P}_{post} = \mathcal{P}_{complete} = 350 ns \cdot k$, and 
$\mathcal{P}_{start} = 0.7 \mu s$, $\mathcal{P}_{wait} = 1.8 \mu s$,

\paragraph{Passive Target Synchronization}

The performance of lock/unlock is constant in the number of processes
(due to the global/local locking) and thus not graphed. The performance
functions are
$\mathcal{P}_{lock,excl} = 5.4\mu s$, $\mathcal{P}_{lock,shrd} =
\mathcal{P}_{lock\_all} = 2.7\mu s$, 
$\mathcal{P}_{unlock} =  \mathcal{P}_{unlock\_all} = 0.4 \mu s$, 
$\mathcal{P}_{flush} = 76 ns$, and $\mathcal{P}_{sync} = 17 ns$.

We demonstrated the performance of our protocols and implementation
using microbenchmarks comparing to other RMA and message passing
implementations. The exact performance models for each call can be
utilized to design and optimize parallel applications, however, this is
outside the scope of this paper. To demonstrate the usability and
performance of our protocols for real applications, we continue with a
large-scale application study. 

\section{Application Evaluation}

We selected two motif applications to compare our protocols and
implementation with the state of the art: a distributed
hashtable representing many big data and analytics applications and a
dynamic sparse data exchange representing complex modern scientific
applications. We also analyze the application MIMD Lattice Computation
(MILC), a full production code with several hundred thousand source
lines of code, as well as a 3D FFT code.

In all codes, we tried to keep most parameters constant to compare the
performance of PGAS languages, MPI-1, and MPI-3.0 RMA. Thus, we did not
employ advanced concepts, such as MPI datatypes or process topologies,
that are not available in all implementations (e.g., UPC and Fortran
Coarrays). 

\subsection{Distributed Hashtable}

Our simple hashtable represents data analytics applications that
often require random access in distributed structures. We compare
MPI point-to-point communication, UPC, and MPI-3.0 RMA.  
In the implementation, each process manages a part of the hashtable
called the \emph{local volume} consisting of a table of elements and an
additional overflow heap to store elements after collisions. The table
and the heap are constructed using fixed-size arrays. In order to avoid
traversing of the arrays, pointers to most recently inserted items as
well as to the next free cells are stored along with the remaining data
in each local volume. The elements of the hashtable are 8-Byte
integers.

The MPI-1 implementation is based on MPI Send and Recv using an
\emph{active message} scheme. Each
process that is going to perform a remote operation sends the element to
be inserted to the owner process which invokes a handle to
process the message. Termination detection is performed using a simple
protocol where each process notifies all other processes of its local
termination.  
In UPC, table and overflow list are placed in shared arrays. Inserts are
based on proprietary (Cray--specific extensions) atomic compare and
swap (CAS) operations. If a collision happens, the losing thread
acquires a new element in the overflow list by atomically incrementing
the next free pointer. It also updates the last pointer using a second
CAS. UPC\_fences are used to ensure memory consistency.
The MPI-3.0 implementation is rather similar to the UPC implementation,
however, it uses MPI-3.0's standard atomic operations combined with
flushes.

Figure~\ref{fig:hashtablePlots} shows the inserts per second for a batch
of 16k operations per process, each adding an element to a random key
(which resides at a random process). MPI-1's performance is competitive for
intra-node communications but inter-node overheads significantly impact
performance and the insert rate of a single
node cannot be achieved with even 32k cores (optimizations such as
coalescing or message routing and reductions~\cite{active-pebbles} would improve
this rate but significantly complicate the code). \fompi{} and UPC
exhibit similar performance characteristics with \fompi{} being slightly
faster for shared memory accesses. The spikes at 4k and 16k nodes are
caused by different job layouts in the Gemini torus and different
network congestion.

\subsection{Dynamic Sparse Data Exchange}

The dynamic sparse data exchange (DSDE) represents a common pattern in
irregular applications~\cite{hoefler-dsde}. DSDE is used when a
set of senders has data destined to arbitrary target processes but no
process knows the volume or sources of data it needs to receive. 
The DSDE pattern is very common in graph-based computations, n-body
simulations, and adaptive mesh refinement codes. Due to the lack of
space, we use a DSDE microbenchmark as proxy for the communication
performance of such applications~\cite{hoefler-dsde}. 

In the DSDE benchmark, each process picks $k$ targets randomly and
attempts to send eight Bytes to each target. The DSDE protocol can
either be implemented using alltoall, reduce\_scatter, a nonblocking
barrier combined with synchronous sends, or one sided accumulates in active
target mode. The algorithmic details of the protocols are described
in~\cite{hoefler-dsde}.
Here, we compare all protocols of this application microbenchmark with
the Cray MPI-2.2 and \fompi{} MPI-3.0 implementations.
Figure~\ref{fig:applications:dsde}
shows the times for the complete exchange using the four different
protocols (the accumulate protocol is tested with Cray's MPI-2.2
implementation and \fompi{}) and $k=6$ random neighbors per process. The
RMA-based implementation is competitive with the nonblocking barrier,
which was proved optimal in~\cite{hoefler-dsde}. \fompi{}'s accumulates
have been tuned for Cray systems while the nonblocking barrier we use is
a generic dissemination algorithm. The performance improvement relative
to other protocols is always significant and varies between a factor of
two and nearly two orders of magnitude.

\subsection{3D Fast Fourier Transform}

We now discuss how to exploit overlap of computation and communication
with our low-overhead implementation in a three-dimensional Fast Fourier
Transformation. We use the MPI and UPC versions of the NAS 3D FFT
benchmark.  Nishtala et al.~and Bell et
al.~\cite{Nishtala:2009:SCA:1586640.1587648,Bell:2006:OBL:1898953.1899016}
demonstrated that overlap of computation and communication can be used
to improve the performance of a 2D-decomposed 3D FFT. We compare the
default ``nonblocking MPI'' with the ``UPC slab'' decomposition, which
starts to communicate the data of a plane as soon as it is available and
completes the communication as late as possible. For a fair comparison,
our \fompi{} implementation uses the same decomposition and
communication scheme like the UPC version and required minimal code
changes resulting in the same code complexity. 

Figure~\ref{fig:applications:fft} shows the performance for the strong
scaling class D benchmark ($2048\times 1024\times 1024$) on different core
counts. UPC achieves a consistent speedup over MPI-1, mostly due to the
overlap of communication and computation. \fompi{} has a slightly lower
static overhead than UPC and thus enables better overlap
(cf.~Figure~\ref{fig:microbench:overlap}), resulting in a slightly
better performance of the FFT.

\subsection{MIMD Lattice Computation}

The MIMD Lattice Computation (MILC) Collaboration studies Quantum
Chromodynamics (QCD), the theory of strong interaction
\cite{bernard91}. The group develops a set of applications, known as
the MILC code. In this work, we use version 7.6.3 as a base. That code
regularly gets one of the largest allocations of computer time at US NSF
supercomputer centers.  The su3\_rmd code, which is part of the SPEC
CPU2006 and SPEC MPI benchmarks, is included in the MILC code.

The code performs a stencil computation on a four-dimensional
rectangular grid.  Domain decomposition is performed in all four
dimensions to minimize the surface-to-volume ratio. In order to keep
data consistent, neighbor communication is performed in all eight
directions, in addition, global allreductions are done regularly to
check convergence of a solver. The most time consuming part of MILC is
the conjugate gradient solver which uses nonblocking MPI
communication overlapped with local computations. 

The performance of the full code and the solver have been analyzed in
detail in~\cite{milc-modeling}.  Several optimizations have been
applied, and a UPC version that demonstrated significant speedups is
available~\cite{upcmilc}. This version replaces the MPI communication
with a simple remote memory access protocol. A process notifies all
neighbors with a separate atomic add as soon as the data in the ``send''
buffer is initialized. Then all processes wait for this flag before they
get (using Cray's proprietary \verb~upc_memget_nb~) the communication
data into their local buffers. This implementation serializes the data from
the application buffer into UPC's communication buffers.
Our MPI-3.0 implementation follows the same scheme to ensure a fair
comparison. We place the
communication buffers into MPI windows and use \mpi{Fetch\_and\_op} and
\mpi{Get} with a single lock all epoch and \mpi{Win\_flush} to perform the
communications. The necessary changes are small and the total number of
source code lines is equivalent to the UPC version. We remark that
additional optimizations may be possible with MPI, for example, one
could use MPI datatypes to communicate the data directly from the
application buffers resulting in additional performance
gains~\cite{hoefler-datatypes}. However, since our goal is to compare to
the UPC version, we only investigate the packed version here.

\begin{figure}
  \center
  \includegraphics[width=0.85\columnwidth]{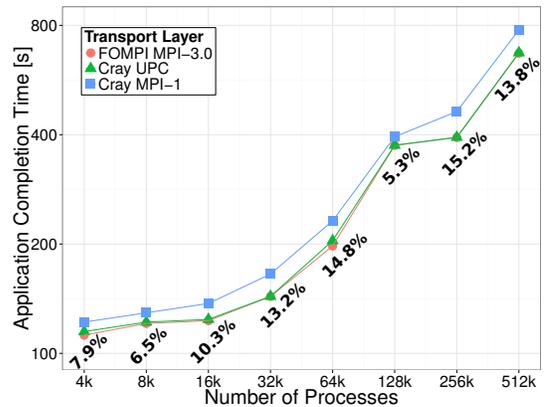}
  \caption{MILC: Full application execution time. The annotations represent the
   improvement of \fompi{} and UPC over MPI-1.}
  \label{fig:applications:milc}
\end{figure}

Figure~\ref{fig:applications:milc} shows the execution time of the whole
application for a weak-scaling problem with a local lattice of $4^3
\times 8$, a size very similar to the original Blue Waters Petascale
benchmark. Some phases (e.g., CG) of the computation execute up to 45\%
faster, however, we chose to report full-application performance. 
The UPC and \fompi{} codes exhibit essentially the same performance,
while the UPC code uses Cray-specific tuning and the MPI-3.0 code is
portable to different architectures. The full-application performance
gain over the MPI-1 version is more than 15\% for some configurations.
The application was scaled successfully up to 524,288 processes with all
implementations.
This result and our microbenchmark demonstrate the scalability and
performance of our protocols and that the new RMA semantics can be used
to improve full applications to achieve performance close to the
hardware limitations in a fully portable way. Since most of those existing
applications are written in MPI, a step-wise transformation can be used
to optimize most critical parts first. 

\section{Related Work}

The intricacies of MPI-2.2 RMA implementations over InfiniBand networks
have been discussed by Jian et al. and Santhanaraman et
al.~\cite{Jiang:2004:HPM:1111683.1111870,Santhanaraman:2009:NST:1577849.1577927}. 
Zhao et al. describe an adaptive strategy to switch from eager to lazy
modes in active target synchronizations in MPICH
2~\cite{Zhao:2012:ASO:2404033.2404043}. This mode could be used to speed
up \fompi{}'s atomics that are not supported in hardware.

PGAS programming has been investigated in the context of UPC and Fortran
Coarrays.
An optimized UPC Barnes Hut implementation shows similarities to MPI-3.0
RMA programming by using bulk vectorized memory transfers combined with
vector reductions instead of shared pointer
accesses~\cite{Zhang:2011:OBA:2063384.2063485}. Nishtala at al.~and Bell
et al.~used overlapping and one sided accesses to improve FFT
performance~\cite{Nishtala:2009:SCA:1586640.1587648,Bell:2006:OBL:1898953.1899016}.
Highly optimized PGAS applications often use a style that can easily be
adapted to MPI-3.0 RMA.

The applicability of MPI-2.2 One Sided has also been demonstrated for some
applications. 
Mirin et al. discuss the usage of MPI-2.2 One Sided coupled with threading
to improve the Community Atmosphere Model (CAM)
\cite{Mirin:2005:SIF:1093627.1093636}. Potluri et al.~show that MPI-2.2
One Sided with overlap can improve the communication in a Seismic
Modeling application \cite{Potluri:2010:QPB:1810085.1810092}. However,
we demonstrated new MPI-3.0 features that can be used to further improve
performance and simplify implementations.

\section{Discussion and Conclusions}

In this work, we demonstrated how MPI-3.0 can be implemented over RDMA
networks to achieve similar performance to UPC and Fortran Coarrays while offering 
all of MPI's convenient functionality (e.g., Topologies and Datatypes). 
We provide detailed performance models, that help choosing among the
multiple options. For example, a user can use our models to decide
whether to use Fence or PSCW synchronization (if $\mathcal{P}_{fence} >
\mathcal{P}_{post}+\mathcal{P}_{complete}+\mathcal{P}_{start}+\mathcal{P}_{wait}$,
which is true for large $k$). This is just one example for the possible
uses of our detailed performance models.

We studied all overheads in detail and provide instruction counts for
all critical synchronization and communication functions, showing that
the MPI interface adds merely between 150 and 200 instructions in the
fast path. This demonstrates that a library interface like MPI is
competitive with compiled languages such as UPC and Fortran Coarrays.
Our implementation proved to be scalable and robust while running on
524,288 processes on Blue Waters speeding up a full application run by
13.8\% and a 3D FFT on 65,536 processes by a factor of two. 

We expect that the principles and extremely scalable synchronization
algorithms developed in this work will act as a blue print for optimized
MPI-3.0 RMA implementations over future large-scale RDMA networks. We also
expect that the demonstration of highest performance to users will
quickly increase the number of MPI RMA programs. 

{\small 
\subsubsection*{Acknowledgments}
We thank Timo Schneider for early help in the project, Greg Bauer and
Bill Kramer for support with Blue Waters, Cray's Duncan Roweth and
Roberto Ansaloni for help with Cray's PGAS environment, Nick
Wright for the UPC version of MILC, and Paul Hargrove for
the UPC version of NAS-FT. 
This work was supported in part by the DOE Office of Science, Advanced
Scientific Computing Research, under award number DE-FC02-10ER26011,
program manager Lucy Nowell. This work is partially supported by the
Blue Waters sustained-petascale computing project, which is supported by
the National Science Foundation (award number OCI 07-25070) and the
state of Illinois.
}

\bibliographystyle{abbrv}
\bibliography{sc_fompi}

\end{document}